\documentclass[12pt, fleqn]{article}
\usepackage[cp1251]{inputenc}
\usepackage{latexsym,amsfonts,amssymb}
\usepackage{graphicx}

\usepackage{amsbsy}
\usepackage{amsmath}
\usepackage{epsf}
\usepackage{cite}

\newcommand{\plussupset}{%
  \mathrel{%
    \ooalign{%
      $\supset$\cr
      \hidewidth\kern-0.1em\raisebox{0.09ex}{$\scriptstyle+$}\hidewidth\cr
    }%
  }%
}

\newtheorem{theorem}{Theorem}

\newtheorem{remark}{Remark}
\newcommand{\bt}{\begin{theo}}
\newcommand{\et}{\end{theo}}
\newcommand{\bd}{\begin{displaymath}}
\newcommand{\ed}{\end{displaymath}}

\newcommand{\be} {\begin{equation}}
\newcommand{\ee} {\end{equation}}
\newcommand{\ba} {\begin{array}{l}}
\newcommand{\ea} {\end{array}}
\newcommand{\bea}{\begin{eqnarray}}
\newcommand{\eea} {\end{eqnarray}}

\newcommand{\p} {\partial}

\begin{document}

\begin{center}
 {\Large \bf
Lie symmetry classification and exact solutions \\ of a diffusive
Lotka--Volterra  system with convection }

\medskip

{\bf Roman Cherniha $^{a,b,}$\footnote{\small  Corresponding author.
E-mail: r.m.cherniha@gmail.com; roman.cherniha1@nottingham.ac.uk}
and   Vasyl' Davydovych$^{c}$ }

$^{a}$ \quad School of Mathematical Sciences, University of Nottingham,\\
  University Park, Nottingham NG7 2RD, UK

$^{b}$  \quad National University of Kyiv-Mohyla Academy, 2,
Skovoroda Street, Kyiv  04070, Ukraine.

$^{c}$ \quad Institute of Mathematics,  National Academy of Sciences
of Ukraine, \\
 3, Tereshchenkivs'ka Street, Kyiv 01004, Ukraine.\\

\end{center}

\begin{abstract}
A mathematical model  for description  of the viscous fingering
 induced by a  chemical   reaction is under study. This complicated
 five-component model  is reduced to a
three-component diffusive Lotka--Volterra    system with convection
by introducing  a stream function. The system obtained is examined
by the classical Lie method. A complete Lie symmetry classification
is derived via a rigorous algorithm. In particular, it is proved
that the widest Lie algebras of invariance occur when the stream
function generate a linear   velocity field.  The most interesting
cases (from the  symmetry and applicability point of view) are
further studied in order to derive exact solutions. A wide range of
exact solutions are constructed for radially-symmetric stream
functions. These solutions include time-dependent and radially
symmetric solutions as well as more complicated solutions expressed
in terms of the Weierstrass function.
It was shown that some of  exact  solutions can be used for
demonstration of  spatiotemporal evolution of concentrations
corresponding to two reactants  and  their product.

\end{abstract}

\section{Introduction } \label{sec-1}

In \cite{ge-de_wit-2009}, a remarkable  mathematical model is
introduced for description  of the viscous fingering
 induced by the  chemical   reaction of the standard form  $A+B \to C$.
The model reads as
\begin{equation}\label{0-1}\begin{array}{l}
\nabla\cdot U=0, \medskip\\
\kappa \nabla p+\mu(w)U=0, \medskip\\
u_t+ U\cdot\nabla u = d_1 \triangle u-kuv, \medskip\\
v_t+ U\cdot\nabla v = d_2 \triangle v-kuv, \medskip\\
w_t+U\cdot\nabla w = d_3 \triangle w+kuv, \medskip\\
\end{array}\end{equation}
where the operator $\nabla$ and the Laplacian $\triangle$ are taken
in $\bf{R}^2$. The functions and coefficients in  (\ref{0-1})  have
the following physical meanings: the functions $u(t,x,y), \
v(t,x,y)$ and $w(t,x,y)$ denote two reactants $A$ and $B$ and their
product $C$, respectively; $k$ is a kinetic constant; $p(t,x,y)$ is
pressure; $U=(U_1,U_2)$ is  two-dimensional velocity field; $d_1, \
d_2$ and $d_3$ are  diffusion coefficients; $\kappa$ is
permeability; $\mu(w)$ is  viscosity of the fluid.

 The nonlinear  model (\ref{0-1}) was further
studied by different mathematical techniques in many papers, for
example, in recent studies
\cite{trevelyan-2018,sharma-2019,wit-2020,omori-2020,verma-2022}.
However, to the best of our knowledge, this model was not examined
by the symmetry-based methods and its exact solutions are unknown at
the present time. Because there is no existing general theory for
integration of nonlinear partial differential equations (PDEs),
construction of particular exact solutions for these equations
remains an important mathematical  problem. Finding exact solutions
that have a clear interpretation for the given process is of
fundamental importance. Notably, in contrast to linear PDEs, the
well-known principle of linear superposition cannot be applied to
generate new exact solutions for nonlinear PDEs. Nowadays,  the most
powerful  methods for construction of exact solutions to nonlinear
PDEs are the symmetry-based methods, in particular  the Lie method
and the method of conditional (including nonclassical)  symmetries.
There are thousands of papers devoted to the application of
symmetry-based methods to PDEs; therefore, we list only several
recent monographs, such as \cite{bl-anco-10, arrigo-15, ch-da-2017,
ch-se-pl-2018}.

It should be pointed out  that typically the authors examine scalar
PDEs because  search for symmetries of  {\it systems of nonlinear
PDEs} is a much complicated problem. In fact, essential technical
difficulties occur if one intends to identify symmetries and
construct exact solutions for  systems of PDEs. A typical example is
the very recent study \cite{mandal-2025}, in which the authors for
the very beginning  consider  a three-component system arising in
fluid dynamics. However, in order to identify symmetries, they
simplify the system to  a single fourth-order PDE and, even have
done this, they were able to obtain only particular  results about
symmetries of the PDE in question \cite{ch-arxiv-2025}.  There are
some studies devoted to application of symmetry-based methods to
systems of nonlinear  PDEs, in particular, the papers devoted to
{\it multicomponent systems of PDEs} (system (\ref{0-1}) is
multicomponent because consists of more than two PDEs).
   Taking into account the above observation, we  refer  the reader to the recent works
 \cite{ch-dav-2019,ch-dav-2020,nadjaf-2021,ch-dav-2022,pal-2023,shag-2023,ch-dav-vor-2024,oliveri-25}, devoted to applications of symmetry-based
 methods  to the nonlinear  multicomponent systems of PDEs.

In this work, we introduce a stream function  according to the
well-known formulae (actually, it was done in \cite{ge-de_wit-2009}
as well) and immediately obtain a four-component system with a
semiautonomous equation for the pressure $p(t,x,y)$. As a result, a
three-component diffusive Lotka--Volterra (DLV)   system with
convection terms  is examined instead of the five-component model
(\ref{0-1}). The rest of this paper  is organized as follows. In
Section~\ref{sec-2}, a {\it complete Lie symmetry classification}
(LSC) of the derived DLV type system is presented. It should be
stressed that  typically the complete LSC is a highly nontrivial
problem (see Chapter 2 in \cite{ch-se-pl-2018} in detail).
Especially, this problem is difficult if the coefficients of the PDE
system in question  are prescribed as arbitrary functions of two or
more variables.
As a result, there are many studies in which instead of the complete
LSC  only  particular cases of Lie symmetry of a given system are
identified.

In Section~\ref{sec-3}, two most interesting cases of the  DLV type
system, which follow from the LSC obtained in Section~\ref{sec-2},
are examined in order to construct multiparameter families of exact
solutions. Both cases correspond to the  radially-symmetric  stream
function $\Psi$, which naturally arises in real-world applications
\cite{batchelor-67,tryggvason-83,tan-87}. In particular, we examine
in detail the simplest case when $\Psi=x^2+y^2$ because a very reach
Lie symmetry occurs for this stream function. An analysis is
performed in order to  show that relevant Lie symmetries  form a
highly unusual representation of a well-known five-dimensional Lie
algebra. The latter occurs, for example, for the standard nonlinear
diffusion equation. Having done the above analysis, the well-known
technique based on reduction of the PDE system in question to
ordinary differential equations  was applied for finding exact
solutions. Plots of a family of the solutions derived have been
drawn to show their properties. Finally, we discuss the results
obtained and present some conclusions  in Section~\ref{sec-4}.

\section{Lie symmetry classification } \label{sec-2}
Because the pressure $p$ arises only in the second equation of
(\ref{0-1}),  this equation can easily be solved at the final stage
when the velocity vector $U$ and the concentration $w$ are derived
from other equations. Moreover, the first equation can be
automatically satisfied if one introduces the stream function $\Psi$
according to the well-known formulae\,: $U_1=\frac{\p \Psi}{\p y}$
and $U_2=-\frac{\p \Psi}{\p x}.$ We also assume that the space
derivatives of the stream function do not depend on time, i.e. they
are the functions of $x$ and $y$ only. As a result, we obtain the
three-component   system
\begin{equation}\label{2-1}\begin{array}{l}
u_t+\Psi_yu_x-\Psi_xu_y = d_1\left(u_{xx}+u_{yy}\right)-kuv, \medskip\\
v_t+\Psi_yv_x-\Psi_xv_y = d_2\left(v_{xx}+v_{yy}\right)-kuv, \medskip\\
w_t+\Psi_yw_x-\Psi_xw_y = d_3\left(w_{xx}+w_{yy}\right)+kuv. \medskip\\
\end{array}\end{equation}

 It can be noted that the first two equations in
(\ref{2-1}) form a diffusive Lotka--Volterra system with convective
terms. Actually, the third equation has the very similar structure,
however the quadratic term $kuv$ does not involve $w$. So, we refer
to (\ref{2-1}) as the diffusive Lotka--Volterra system (DLVS) with
convection in what follows.

Our aim is   solving the LSC problem for system (\ref{2-1}). It
means that one should identify all possible forms of the function
$\Psi(x,y)$ leading to extensions of a so-called principal algebra.
According to the definition,
 the  principal algebra  is  derived  under assumption that $\Psi(x,y)$ is an  arbitrary function.
 The detailed algorithm for solving LSC problem for a given PDE (system of PDEs) involving  arbitrary function(s)
 as parameter(s) is described  in \cite[Chapter 2]{ch-se-pl-2018}.
 Of course, this algorithm can be modified depending on the form of
 equation(s) in question. The first step usually consists of finding the group of equivalence
transformations (ETs). So,
  we present a statement about  ETs  of system  (\ref{2-1}).

\begin{theorem} \label{th1} System (\ref{2-1}) can be transformed into a system
of the same structure
\begin{equation}\label{2-2}\begin{array}{l}
u^*_{t^*}+\Psi^*_{y^*}u^*_{x^*}-\Psi^*_{x^*}u^*_{y^*} = d^*_1\left(u^*_{x^*x^*}+u^*_{y^*y^*}\right)-k^*u^*v^*, \medskip\\
v^*_{t^*}+\Psi^*_{y^*}v^*_{x^*}-\Psi^*_{x^*}v^*_{y^*} = d^*_2\left(v^*_{x^*x^*}+v^*_{y^*y^*}\right)-k^*u^*v^*, \medskip\\
w^*_{t^*}+\Psi^*_{y^*}w^*_{x^*}-\Psi^*_{x^*}w^*_{y^*} = d^*_3\left(w^*_{x^*x^*}+w^*_{y^*y^*}\right)+k^*u^*v^*, \medskip\\
\end{array}\end{equation} using equivalence transformations
 \be\label{2-4}\begin{array}{l} t^*=\alpha_0t+t_0, \ x^*=\alpha_1x+\alpha_2y+x_0, \ y^*=\mp\alpha_2x\pm\alpha_1y+y_0, \medskip\\
u^*=\alpha_3 u, \ v^*=\alpha_3 v, \ w^*=\alpha_3 w+H(t,x,y), \medskip\\
d^*_1=\frac{\alpha_1^2+\alpha_2^2}{\alpha_0}\,d_1, \
d^*_2=\frac{\alpha_1^2+\alpha_2^2}{\alpha_0}\,d_2, \
d^*_3=\frac{\alpha_1^2+\alpha_2^2}{\alpha_0}\,d_3, \
k^*=\frac{k}{\alpha_0\alpha_3}, \
\Psi^*=\pm\frac{\alpha_1^2+\alpha_2^2}{\alpha_0}\Psi+\Psi_0,
 \ea\ee
and/or \be\label{2-5}u^*=v, \ v^*=u, \ d^*_1=d_2, \ d^*_2=d_1, \
d^*_3=d_3, \ k^*=k, \ \Psi^*=\Psi,\ee
 where $\alpha_0>0, \ \alpha_1, \ \alpha_2, \ t_0, \ x_0, \ y_0$
and $\alpha_3>0$   are the real group parameters, $H(t,x,y)$ is an
arbitrary solution of the linear equation
\be\label{2-3}H_t+\Psi_yH_x-\Psi_xH_y =
d_3\left(H_{xx}+H_{yy}\right).\ee
\end{theorem}

\textbf{Proof.} Typically,   the known technique based on the
classical Lie method  for constructing the group of continuous ETs
is used.  Because this technique is  cumbersome, there are not many
papers,
 in which it was described in detail and  successfully employed
 for nontrivial PDEs. One of the first examples for two-dimensional PDEs was presented  in \cite{ibr-tor-1991}
 (see also a recent study  \cite{ch-da-QTDS-25} for multidimensional
 PDEs). Here  a so-called  direct method was employed to construct ETs.
  The  direct method requires to start from the most general form of
  point transformations for the given equation(s). In the case of system
  (\ref{2-1}), one should start from the transformations\,: \be\label{2-4*}\begin{array}{l} t^*=f(t,x,y,u,v,w), \
x^*=g(t,x,y,u,v,w), \ y^*=h(t,x,y,u,v,w),\\
u^*=F(t,x,y,u,v,w), \ v^*=G(t,x,y,u,v,w), \ w^*=H(t,x,y,u,v,w),
\end{array}\end{equation}
where $f, \ g, \ h, \ F, \ G$ and $H$ are arbitrary smooth functions
with nonvanishing Jacobian
\begin{equation}\nonumber
\det \frac{\partial(t^*,x^*,y^*,u^*,v^*,w^*)}{\partial(t,x,y,u,v,w)}
\neq 0.
\end{equation}
According to the definition of ETs, one should find all possible
point transformations of the form (\ref{2-4*}),
which transform  system  (\ref{2-1}) with an arbitrary function
$\Psi$  into a system with  the same structure  involving  a
function $\Psi^*$ that can be different from $\Psi$. Generally
speaking, relevant calculations are very cumbersome (see a detailed
example in \cite[Chapter 2]{ch-se-pl-2018}). However, it can be
easily shown that transformations for  independent variables  are
essentially simplified in the case of system (\ref{2-1}), namely:
\be\nonumber t^*=f(t), \ x^*=g(x,y), \ y^*=h(x,y), \ f'(t)\neq0, \
\frac{\partial(g,h)}{\partial(x,y)} \neq 0.\ee As a result, formulae
(\ref{2-4})  were derived using  straightforward calculations. \
$\blacksquare$

 Formally speaking, the equivalence transformations
(\ref{2-4}) form an infinite-parameter Lie group. However, the
function $H$  and the linear PDE (\ref{2-3}) reflect an obvious fact
that the third equation in (\ref{2-1}) is linear with respect to $w$
 and the first two equation do not depend explicitly on $w$,
  therefore we will not pay special attention to this in what
follows.
\begin{theorem} \label{th2}
System (\ref{2-1}) with an arbitrary  function $\Psi(x,y)$ and
arbitrary positive coefficients $k$ and $d_i \ (i=1,2,3)$ is
invariant under
the principal algebra with the basic operators \be\label{p-6}\p_t, \
H(t,x,y)\p_w,\ee where $H$ is an
arbitrary solution of the linear equation (\ref{2-3}).\\

 In the
special cases $d_1=d_3, \ d_2\neq d_3$ and $d_2=d_3, \ d_1\neq d_3$,
the principal algebra additionally involves the Lie symmetry
operator $(u+w)\p_w$ and $(v+w)\p_w$, respectively. Both above
operators occur for   system (\ref{2-1}) with $d_1=d_2=d_3$.
\end{theorem}

\begin{theorem} \label{th3}  The DLVS with
convection  (\ref{2-1}), depending on the function $\Psi(x,y)$,
admits exactly 11 extensions of the principal algebra, which  are
listed in Table~\ref{tab1}. Any system
 (\ref{2-1})  with different form of  $\Psi(x,y)$ is either invariant w.r.t. the principal
algebra, or is reducible to one listed in Table~\ref{tab1} by the
ETs (\ref{2-4}).
\end{theorem}
\begin{small}
\begin{table}[h!]
\caption{Lie symmetries of system (\ref{2-1})}\medskip
\label{tab1}       
\begin{tabular}{|c|c|c|}
\hline  & Restrictions  &  Additional Lie symmetries  \\
 \hline && \\
  1 & $\Psi=F\left(x^2+y^2\right)+\left(\alpha+\beta\left(x^2+y^2\right)\right)\arctan\left(\frac{x}{y}\right)$  &
  $e^{2\beta t}\left(y\p_x-x\p_y\right)$ \\ \hline &&\\
 2 & $\Psi=F\left(\alpha_1x+\alpha_2y\right)+\beta x\left(\alpha_1x+\alpha_2y\right)+\gamma x$  &
  $e^{\alpha_2\beta t}\left(\alpha_2\p_x-\alpha_1\p_y\right)$ \\
 \hline &&\\
 3 & $\Psi=F\left(\frac{x}{y}\right)+\alpha \ln x$  & $2t\p_t+x\p_x+y\p_y-2u\p_u-2v\p_v-2w\p_w$ \\
  \hline &&\\
  4 & $\Psi=F\left(\arctan\left(\frac{x}{y}\right)+
  \alpha_0\ln\left(x^2+y^2\right)\right)+$  & $2t\p_t+\left(x-2\alpha_0 y\right)\p_x+$
  \\ &$\alpha\arctan\left(\frac{x}{y}\right)$&$\left(y+2\alpha_0 x\right)\p_y-2u\p_u-2v\p_v-2w\p_w$ \\
  \hline &&\\
  5 & $\Psi=\alpha\arctan\left(\frac{x}{y}\right)+\beta\ln\left(x^2+y^2\right)+\gamma\left(x^2+y^2\right)$  &
  $y\p_x-x\p_y, \ 2t\p_t+\left(x+4\gamma ty\right)\p_x+$ \\ &$\alpha^2+\beta^2+\gamma^2\neq0$&$\left(y-4\gamma tx\right)\p_y-2u\p_u-2v\p_v-2w\p_w$ \\
  \hline &&\\
   6 & $\Psi=\pm
\ln\left(\alpha_1x+\alpha_2y\right)+\gamma\left(\alpha_1x+\alpha_2y\right)$
& $\alpha_2\p_x-\alpha_1\p_y, \
2t\p_t+\left(x+\gamma\alpha_2t\right)\p_x+$ \\ &$\alpha_1^2+\alpha_2^2\neq0$&$\left(y-\gamma\alpha_1t\right)\p_y-2u\p_u-2v\p_v-2w\p_w$\\
  \hline &&\\
   7 & $\Psi=\alpha_0 x^2+\frac{y^2}{4\alpha_0}, \ \alpha_0\neq\pm\frac{1}{2}$  &
    $\cos t\,\p_x-2\alpha_0\sin t\,\p_y, \ \sin t\,\p_x+2\alpha_0 \cos t\,\p_y$
    \\
  \hline &&\\
  8 & $\Psi=\alpha_0 x^2-\frac{y^2}{4\alpha_0}$  & $e^t\left(\p_x-2\alpha_0\p_y\right), \ e^{-t}\left(\p_x+2\alpha_0\p_y\right)$ \\
  \hline &&\\
  9 & $\Psi=x^2+\alpha y$  & $\p_y, \ \p_x-2t\p_y$ \\
  \hline &&\\
  10 & $\Psi=x^2+y^2$  & $y\p_x-x\p_y, \ \sin(2 t)\p_x+\cos(2 t)\p_y,$
  \\ && $\cos(2t)\p_x-\sin(2t)\p_y, \ 2 t\p_t+(x+4ty)\p_x+$ \\
  && $(y-4tx)\p_y-2 u\p_u-2 v\p_v-2 w\p_w$ \\
  \hline &&\\
   11 & $\Psi=\alpha_1x+\alpha_2y$  &
$\p_x, \ \p_y,  \ 2t\p_t+(x+\alpha_2t)\p_x+$ \\
&& $(y-\alpha_1t)\p_y-2u\p_u-2v\p_v-2w\p_w$\\
&&$(\alpha_1t+y)\p_x+(\alpha_2t-x)\p_y$\\
   \hline
 \end{tabular} \medskip

In Table~\ref{tab1}, $\alpha_0\neq0, \ \alpha_1, \ \alpha_2,\
\alpha, \ \beta$ and $\gamma$ are arbitrary constants, $F$ is an
arbitrary smooth function.
\end{table}
\end{small}

\begin{remark} Some functions $\Psi$ presented in Table~\ref{tab1} can be further simplified using
 form-preserving (admissible)  transformations introduced independently in \cite{gaz-wint-92} and  \cite{kingston-98} for classification
 of PDEs. The transformation $x^*=x-\gamma\alpha_2t, \
y^*=y+\gamma\alpha_1t$ makes  $\gamma=0$ in Case 6; transformation
$x^*=x-\alpha_2t, \ y^*=y+\alpha_1t$  makes  $\alpha_1=\alpha_2=0$
in Case 11.
\end{remark}

\begin{remark} The classical example of the velocity field,
 a vortex flow around the origin $(U_1,U_2)=\Big(\frac{\beta y}{x^2+y^2}, \frac{-\beta
 x}{x^2+y^2}\Big)$, can be identified in Case 5 with
 $\alpha=\gamma=0$. For the vortex flow, system (\ref{2-1}) admits
 two additional Lie symmetries corresponding to rotations and scale
 transformations.
 \end{remark}

\textbf{Proof of Theorems~\ref{th2} and \ref{th3}.} Here we use the
algorithm, which was suggested in  \cite[Chapter 2]{ch-se-pl-2018}
for LSC of evolutionary equations. Because the group of ETs is
already identified, we need to find the principal algebra at the
next step.

So, we start from to the most general form of Lie symmetries of
system (\ref{2-1}):
 \be \nonumber \ba
X=\xi^0\left(t,x,y,u,v,w\right)\partial_t+\xi^{1}\left(t,x,y,u,v,w\right)\partial_{x}+\xi^{2}\left(t,x,y,u,v,w\right)\partial_{y}+\medskip\\
\hskip2cm
\eta^1\left(t,x,y,u,v,w\right)\partial_{u}+\eta^2\left(t,x,y,u,v,w\right)\partial_{v}+\eta^3\left(t,x,y,u,v,w\right)\partial_{w},\ea\ee
where  $\xi^i, \ i=0,1,2$ and  $\eta^k, \ k=1,2,3$ are
to-be-determined  functions. The well-known infinitesimal criterion
of invariance of (\ref{2-1}) with respect to the symmetry $X$ reads
as
\[\ba \mbox{\raisebox{-1.6ex}{$\stackrel{\displaystyle
X}{\scriptstyle 2}$}}\,
\left(d_1\left(u_{xx}+u_{yy}\right)-u_t-\Psi_yu_x+\Psi_xu_y-kuv\right)\Big\vert_{\cal{M}}=0,
\medskip \\
\mbox{\raisebox{-1.6ex}{$\stackrel{\displaystyle X}{\scriptstyle
2}$}}\,
\left(d_2\left(v_{xx}+v_{yy}\right)-v_t-\Psi_yv_x+\Psi_xv_y-kuv\right)\Big\vert_{\cal{M}}=0,
\medskip\\
\mbox{\raisebox{-1.6ex}{$\stackrel{\displaystyle X}{\scriptstyle
2}$}}\,
\left(d_3\left(w_{xx}+w_{yy}\right)-w_t-\Psi_yw_x+\Psi_xw_y+kuv\right)\Big\vert_{\cal{M}}=0,\ea\]
where the operator $
\mbox{\raisebox{-1.6ex}{$\stackrel{\displaystyle X}{\scriptstyle
2}$}}$ is the second-order prolongation of the operator $X$, and the
manifold   ${\cal{M}}$ consists of the equations of the system in
question.

 Using the  above  infinitesimal criterion  and  carrying out  relevant  computations,
 the functions $\xi^i, \ i=0,1,2$ and  $\eta^k, \ k=1,2,3$  were
 specified as follows
\be\nonumber\ba\xi^0=2c_0t+t_0, \ \xi^1=c_0x+p_0(t)y+p_1(t), \
\xi^2=c_0y-p_0(t)x+p_2(t), \medskip\\
\eta^1=-2c_0u, \ \eta^2=-2c_0v, \
\eta^3=c_1u+c_2v+\left(c_1+c_2-2c_0\right)w+H(t,x,y).
 \ea\ee
The constants $c_i$ ($i=0,1,2$), the functions $p_i$ ($i=0,1,2$),
and $H$ should be determined from the system of determining
equations (DEs)
\begin{eqnarray} && (d_1-d_3)\,c_1=0, \ (d_2-d_3)\,c_2=0,\label{p-2} \\
&& H_t+\Psi_yH_x-\Psi_xH_y =
d_3\left(H_{xx}+H_{yy}\right), \label{p-3} \\
&&\left(c_0x+p_0(t)y+p_1(t)\right)\Psi_{xx}+\left(c_0y-p_0(t)x+p_2(t)\right)\Psi_{xy}+\nonumber \\
&& \hskip1cm c_0\Psi_x-p_0(t)\Psi_y-x{p_0}'(t)+{p_2}'(t)=0,\label{p-4} \\
&&\left(c_0y-p_0(t)x+p_2(t)\right)\Psi_{yy}+\left(c_0x+p_0(t)y+p_1(t)\right)\Psi_{xy}+\nonumber \\
&& \hskip1cm
c_0\Psi_y+p_0(t)\Psi_x-y{p_0}'(t)-{p_1}'(t)=0.\label{p-5}
\end{eqnarray}

To identify the principal algebra, one should find necessary and
sufficient conditions when the system of DEs
(\ref{p-2})--(\ref{p-5}) is fulfilled  for an arbitrarily given
function $\Psi(x,y)$ and arbitrary diffusivities $d_i \ (i=1,2,3)$.
So,  one immediately obtains $c_i=p_i=0$ ($i=0,1,2$), therefore only
two Lie symmetry  operators listed in~(\ref{p-6}) are obtained. If
$d_1=d_3$ or $d_2=d_3$, we additionally obtain the operator
$(u+w)\p_w$ or $(v+w)\p_w$, respectively. The case $d_1=d_2=d_3$, of
course, leads to two additional operators. Thus, Theorem~\ref{th2}
is proved.

To obtain the results presented in Table~\ref{tab1}, one must solve
the system of classification equations (\ref{p-4})--(\ref{p-5}) in
such a way that all possible functions $\Psi$, leading to larger Lie
algebras of invariance,  will be specified. Simultaneously,  the
equivalence transformations (\ref{2-4})--(\ref{2-5}) should be
taking into account.

Integrating equations (\ref{p-4})--(\ref{p-5}) with respect to
(w.r.t.) the variables $x$ and $y$, respectively, one obtains two
first-order PDEs. After  a simple analysis, we concluded that both
PDEs obtained are equivalent to the single equation
 \be\label{p-7}
\Big(c_0x+p_0(t)y+p_1(t)\Big)\Psi_{x}+\Big(c_0y-p_0(t)x+p_2(t)\Big)\Psi_{y}-\frac{x^2+y^2}{2}\,{p_0}'(t)+x{p_2}'(t)-y{p_1}'(t)+q(t)=0,
\ee where $q(t)$ is another to-be-determined  function.

The further  analysis is based on  equation (\ref{p-7}). In fact,
the following two cases naturally arise:
 $\textbf{(a)}$ all functions $p_i(t)$ are constants:
 ${p_0}'={p_1}'={p_2}'=0$; $\textbf{(b)}$ at least one function is nonconstant: ${{p_0}'}^2+{{p_1}'}^2+{{p_2}'}^2\neq0.$

\emph{Let us examine Case $\textbf{(a)}$ in  detail}. Equation
(\ref{p-7}) takes the form: \be\label{p-8}
\left(c_0x+p_0y+p_1\right)\Psi_{x}+\left(c_0y-p_0x+p_2\right)\Psi_{y}+q_0=0.
\ee This equation with $c_0=p_0=0$ is a first-order PDE with
constant coefficients, hence \[\Psi=
F(p_2x-p_1y)-q_0\frac{p_1x+p_2y}{p_1^2+p_2^2}\] is its general
solution. The corresponding new symmetry is $p_1\p_x+p_2\p_y$. Thus,
a particular case of Case 2 from Table~\ref{tab1} is identified (up
to notations).

 Equation (\ref{p-8}) with $c_0^2+p_0^2\neq0$
can be rewritten as \be\label{p-9}
\left(c_0x+p_0y\right)\Psi_{x}+\left(c_0y-p_0x\right)\Psi_{y}+q_0=0,
\ee taking into account ET \[x \ \rightarrow
x+\frac{p_0p_2-c_0p_1}{c_0^2+p_0^2}, \ y \ \rightarrow
y-\frac{p_0p_1+c_0p_2}{c_0^2+p_0^2}.\]

Integrating equation (\ref{p-9}) via the classical method of
characteristic, we obtain the function $\Psi$ from Case 3 of
Table~\ref{tab1} provided $c_0\neq0$ and $p_0=0$, and the function
$\Psi$ from Case 4  provided $c_0p_0\neq0$.  Simultaneously,
additional Lie symmetries arising in Cases 3 and 4 are derived.
 Having $c_0=0, \ p_0\neq0$, one obtains the function
$\Psi=F\left(x^2+y^2\right)+\alpha\arctan\left(\frac{x}{y}\right), \
\alpha=-\frac{q_0}{p_0}$, which  appears in Case 1 as a particular
case by setting $\beta=0$. The relevant symmetry is $y\p_x-x\p_y$.

\emph{Analysis of Case $\textbf{(b)}$} is briefly presented below.
One needs to examine two subcases: \[\textbf{(b1)} \ {p_0}'(t)\neq0;
\quad \textbf{(b2)}
 \ {p_0}'(t)=0, \ {{p_1}'}^2+{{p_2}'}^2\neq0\]
 because they lead to different results.

\emph{Subcase $\textbf{(b1)}$}. Because ${p_0}'(t)\not=0$, direct
integration of (\ref{p-7}) leads to a cumbersome expression
therefore we use its differential consequences.   Differentiating
equation (\ref{p-7}) w.r.t. $x, \ y$ and $t$, one obtains  the
equation \be\label{p-13}
y\Psi_{xxy}+\Psi_{xx}-x\Psi_{xyy}-\Psi_{yy}+\frac{{p_1}'}{{p_0}'}\,\Psi_{xxy}+\frac{{p_2}'}{{p_0}'}\,\Psi_{xyy}=0.
\ee  A further  differential consequence of (\ref{p-13}) w.r.t. the
variable $t$ gives  \be\label{p-14} \left({p_0}'{p_1}''-{p_1}'
{p_0}''\right)\Psi_{xxy}+\left({p_0}'{p_2}''-{p_2}'
{p_0}''\right)\Psi_{xyy}=0.\ee Because (\ref{p-14}) is a PDE with
constant coefficients ($t$ plays  role of a parameter), one can be
integrated in a straightforward way. In fact, assuming
${p_0}'{p_1}''-{p_1}' {p_0}''\not=0$, one obtains the equation
\[\Psi_{xxy}+A\Psi_{xyy}=0 \] with the constant coefficient  $ A=
\frac{{p_0}'{p_2}''-{p_2}' {p_0}''}{{p_0}'{p_1}''-{p_1}' {p_0}''}$.
Checking  comparability of the above equation with the generic PDE
(\ref{p-7}), one obtains
$\Psi_{xxy}=\Psi_{xyy}=0$. Thus,  taking into account (\ref{p-13}),
 the most  general form of the function $\Psi$ is
\be\nonumber\Psi(x,y)=\beta\left(x^2+y^2\right)+\alpha_1x+\alpha_2y+\alpha_3xy,\ee
where $\beta$ and $\alpha_i$ are arbitrary constants. Depending on
the values of $\beta$ and $\alpha_i$ and using  the equivalence
transformations (\ref{2-4}),  Cases 9--11 of Table~\ref{tab1} were
identified.

Now we need to examine a special case
${p_0}'{p_1}''-{p_1}' {p_0}''=0$. In this case,   (\ref{p-14})
immediately gives   ${p_0}'{p_2}''-{p_2}' {p_0}''=0$ (assumption
$\Psi_{xyy}=0$  does not lead to new forms of the function $\Psi$).
So, equation (\ref{p-14})  vanishes, while the functions  $p_1$ and
$p_2$  can be derived from the above ODEs: \be\label{p-16}
p_1(t)=\lambda_0+\lambda_1p_0(t), \ p_2(t)=\mu_0+\mu_1p_0(t).\ee
Substituting (\ref{p-16}) into   equation (\ref{p-7}), we arrive at
the equation
\be\nonumber\ba q(t)+\Big((y+\lambda_1)\Psi_{x}-(x-\mu_1)\Psi_{y}\Big)p_0(t)+\left(\mu_1x-\lambda_1y-\frac{x^2+y^2}{2}\right){p_0}'(t)+\medskip\\
\hskip2cm
\left(c_0x+\lambda_0\right)\Psi_{x}+\left(c_0y+\mu_0\right)\Psi_{y}=0
\ea\ee Taking differential consequence  w.r.t. the variable $t$ and
dividing the equation obtained by ${p_0}'(t)\not=0$, we arrive at
\be\label{p-18}
(y+\lambda_1)\Psi_{x}+(\mu_1-x)\Psi_{y}+\beta_1\left(\mu_1x-\lambda_1y-\frac{x^2+y^2}{2}\right)+\beta_0
=0,\ee where \be\nonumber\frac{{p_0}''(t)}{{p_0}'(t)}=\beta_1, \
\frac{q'(t)}{{p_0}'(t)}=\beta_0.\ee Thus, the above ODEs immediately
give  the functions $p_0$ and $q$:
\[p_0(t)=C_1e^{\beta_1t}+C_0, \ q(t)=\beta_0C_1e^{\beta_1t}+C_2\]
if $\beta_1\neq0$, and
\[p_0(t)=C_1t+C_0, \ q(t)=\beta_2C_1t+C_2\]
if $\beta_1=0$.

At the last step, integrating equation (\ref{p-18}), substituting
the obtained functions $\Psi, \ p_i$ and $q$   into (\ref{p-7}), one
obtains algebraic restrictions on arbitrary constants $C_i, \
\beta_j, \ \lambda_j, \ \mu_j$ $(i=0,1,2; \ j=0,1)$.  Finally,
applying the equivalent transformation $x-\mu_1 \ \rightarrow \ x, \
y+\lambda_1 \ \rightarrow \ y$, we arrive at Cases 1 and 5 of
Table~\ref{tab1}.



\emph{Subcase $\textbf{(b2)}$} is much simpler because
differentiation of  equation (\ref{p-7}) w.r.t. $t$ produces the
first-order PDE
\[
p'_1(t)\Psi_{x}+p'_2(t)\Psi_{y}+{p_2}''(t)x-{p_1}''(t)y+q'(t)=0, \]
which can easily be integrated. As a result, Cases 2 and  6--8 from
Table~\ref{tab1} were identified.

\emph{\textbf{The proof is completed.}} $\blacksquare$

\section{Exact solutions} \label{sec-3}

 \subsection{Exact solutions of the  diffusive Lotka--Volterra type system with a specific  stream function} \label{sec-3.2}

 Let us consider the DLVS with
convection corresponding to Case 10 of Table~\ref{tab1}. One notes
that this case represents system (\ref{2-1}) with the linear
velocity field $U=(2y,-2x)$, which  admits the widest Lie algebra of
invariance. Ignoring the Lie symmetry $H(t,x,y)\p_w$, which reflects
linearity of  (\ref{2-1}) w.r.t. the component $w$, the relevant Lie
algebra of invariance is generated by the operators
\begin{equation}\label{4-21} \begin{array}{l}   P_1=\sin(2 t)\p_x+\cos(2 t)\p_y,
  \   P_2=\sin(2t)\p_y-\cos(2t)\p_x, \ J_{12}=y\p_x-x\p_y,
  \medskip\\
  P_t=\p_t, \ D=2 t\p_t+(x+4ty)\p_x+
  (y-4tx)\p_y-2 u\p_u-2 v\p_v-2 w\p_w.
  \end{array} \end{equation}
  It can be easily calculated that the Lie brackets
\begin{equation}\nonumber
[J_{12},P_1]=P_2, \ [J_{12},P_2]=-P_1, \ [P_1,P_2]=0,
 \end{equation}
 therefore we conclude that the Lie algebra $\langle P_1, \ P_2, \
 J_{12}\rangle $ is the well-known Euclid algebra $AE(2)$,  for which
 the standard representation is  $\langle \p_x, \ \p_y, \
 J_{12}\rangle $. The latter generates the tree-dimensional Euclid group of translations
 and rotations in
 the space $\bf{R}^2$.

  Moreover, the four-dimensional Lie algebra (\ref{4-21}) excluding the operator $\p_t$ is
 nothing else but the well-known extension of $AE(2)$ by
 a scaling operator. Typically, it is
  $ D_0=x\p_x+y\p_y$. However,
 (\ref{4-21}) is an unusual representation of the extended Euclid
 algebra $AE(2) \plussupset D_0$: the operator $D$ is used instead of $D_0$.
 In fact, $[D, P_1]=-P_1, \ [D, P_2]=-P_2$ and $[D, J_{12}]=0$.

 Finally, one may consider the five-dimensional Lie algebra (\ref{4-21}) as
  a further extension of the
Euclid algebra $AE(2)$. In fact, the extended Euclid
 algebra $AE_{ext}(2)=P_t\oplus AE(2) \plussupset D_1$ (here $D_1=D_0+2t\p_t$) is a
 Lie algebra  of invariance for many parabolic equations arising in real-world applications.
  A  typical example is the  nonlinear diffusion (heat) equation with an arbitrary diffusivity $D(u)$:
 \[ u_t=\nabla\cdot (D(u)\nabla u).   \]
  In order to satisfy the  Lie
 brackets commutations of the extended Euclid
 algebra $AE_{ext}(2)$, one needs to use the operator $P_t$ in the form
$\p_t+J_{12}$. So, the  Lie algebra (\ref{4-21}) is nothing else but
an unusual representation of $AE_{ext}(2)$.

 Let us construct exact solutions of the system
\begin{equation}\label{2-23}\begin{array}{l}
u_t+2yu_x-2xu_y = d_1\left(u_{xx}+u_{yy}\right)-uv, \medskip\\
v_t+2yv_x-2xv_y = d_2\left(v_{xx}+v_{yy}\right)-uv, \medskip\\
w_t+2yw_x-2xw_y = d_3\left(w_{xx}+w_{yy}\right)+uv,
\end{array}\end{equation}
which corresponds  to Case 10 of Table~\ref{tab1} (we set $k=1$
without loss of generality) and admits  the Lie algebra
(\ref{4-21}). Taking into  account the above analysis of this Lie
algebra, one can reduce the latter to its standard representation
via the transformation \be\label{2-23*}x^*=x\sin 2t+y\cos 2t, \
y^*=y\sin 2t-x\cos 2t.\ee As a result, the basic operators of the
algebra take the form \be\label{3-30*}\langle \p_t, \  \p_x, \ \p_y,
\ J_{12}, \ 2t\p_t+x\p_x+y\p_y-2u\p_u-2v\p_v-2w\p_w \rangle,\ee
while  system (\ref{2-23}) simplifies as follows
\begin{equation}\label{3-30}\begin{array}{l}
u_t = d_1\left(u_{xx}+u_{yy}\right)-uv, \medskip\\
v_t = d_2\left(v_{xx}+v_{yy}\right)-uv, \medskip\\
w_t = d_3\left(w_{xx}+w_{yy}\right)+uv,
\end{array}\end{equation} (hereafter the upper index $*$ is
omitted).

The most general linear combination  of the operators from
(\ref{3-30*}) is given by
 \be\label{3-31} X=(2\alpha t+t_0)\p_t+\left(\alpha
x+\beta y+x_0\right)\p_x+\left(\alpha y-\beta
x+y_0\right)\p_y-2\alpha u\p_u-2\alpha v\p_v-2 \alpha w\p_w.\ee To
construct all  inequivalent  ans\"atze, two essentially different
cases should be examined: \\ \textbf{\emph{(i)}} $\alpha\neq0$ and
\textbf{\emph{(ii)}} $\alpha=0$.
 In
Case \textbf{\emph{(i)}}, operator~(\ref{3-31}) can be simplified
 to the
form \be\label{3-32} X=2\alpha t \p_t+\left(\alpha x+\beta
y\right)\p_x+\left(\alpha y-\beta x\right)\p_y-2\alpha u\p_u-2\alpha
v\p_v-2 \alpha w\p_w\ee
 by
means of the  transformation of independent variables
\[ t \rightarrow t-\frac{t_0}{2\alpha}, \ x \rightarrow \ x+\frac{\beta y_0-\alpha x_0}{\alpha^2+\beta^2}, \ y \
\rightarrow \ y-\frac{\beta x_0+\alpha y_0}{\alpha^2+\beta^2}. \]
Introducing the  the polar coordinates
 \be\nonumber x=r\cos
\varphi, \ y=r\sin \varphi, \ee   system~(\ref{3-30}) and
operator~(\ref{3-32}) are  transformed to the forms
\begin{equation}\nonumber\begin{array}{l}
u_t = d_1u_{rr}+\frac{d_1}{r^2}u_{\varphi\varphi}+\frac{d_1}{r}u_r-uv, \medskip\\
v_t = d_2v_{rr}+\frac{d_2}{r^2}v_{\varphi\varphi}+\frac{d_2}{r}v_r-uv, \medskip\\
w_t =
d_3w_{rr}+\frac{d_3}{r^2}w_{\varphi\varphi}+\frac{d_3}{r}w_r+uv,
\end{array}\end{equation}
and  \be\label{3-35} X=
2\alpha t\p_t+\alpha r\p_r-\beta\p_{\varphi}-2\alpha u\p_u-2\alpha
v\p_v-2 \alpha w\p_w,\ee respectively. Notably,  one may set
$\alpha=1$ in (\ref{3-35}) without loss of generality since
$\alpha\not=0$.
 Obviously,  the corresponding ansatz  can be easily derived and that
reads as follows:
\be\label{3-36}u=\frac{U\left(\omega_1,\omega_2\right)}{t}, \
v=\frac{V\left(\omega_1,\omega_2\right)}{t}, \
w=\frac{W\left(\omega_1,\omega_2\right)}{t}, \
\omega_1=\frac{r^2}{t}, \ \omega_2=\varphi+\beta\ln r.\ee Thus,
using the above ansatz, the following reduced system is obtained
\begin{equation}\label{3-37}\begin{array}{l}
4d_1\omega_1U_{\omega_1\omega_1}+d_1\frac{1+\beta^2}{\omega_1}U_{\omega_2\omega_2}+
4\beta d_1U_{\omega_1\omega_2}+\left(4d_1+\omega_1\right)U_{\omega_1}-UV+U=0, \medskip\\
4d_2\omega_1V_{\omega_1\omega_1}+d_2\frac{1+\beta^2}{\omega_1}V_{\omega_2\omega_2}+
4\beta d_2V_{\omega_1\omega_2}+\left(4d_2+\omega_1\right)V_{\omega_1}-UV+V=0, \medskip\\
4d_3\omega_1W_{\omega_1\omega_1}+d_3\frac{1+\beta^2}{\omega_1}W_{\omega_2\omega_2}+
4\beta
d_3W_{\omega_1\omega_2}+\left(4d_3+\omega_1\right)W_{\omega_1}+UV+W=0.
\end{array}\end{equation}

In Case \textbf{\emph{(ii)}} $\alpha=0$, the operator X (\ref{3-31})
corresponds to the extension of the Euclid algebra $AE(2)$ via the
operator $P_t$. All nonconjugated subalgebras of this algebra can be
found in Table~II of the classical paper \cite{Pat-Wint}. So, the
so-called optimal system of one-dimensional subalgebras consists of
the Lie algebras \be\label{3-43}\langle \p_t\rangle,  \
\langle\p_x+t_0\p_t\rangle, \ \langle  J_{12}+t_0\p_t\rangle. \ee
Obviously, the first algebra from (\ref{3-43}) leads to
time-independent solutions, hence system (\ref{3-30}) reduces to the
form
\begin{equation}\label{3-44}\begin{array}{l}
0 = d_1\left(u_{xx}+u_{yy}\right)-uv, \medskip\\
0 = d_2\left(v_{xx}+v_{yy}\right)-uv, \medskip\\
0 = d_3\left(w_{xx}+w_{yy}\right)+uv.
\end{array}\end{equation}

The second  algebra from (\ref{3-43}) generates the ansatz
\be\nonumber u=U\left(\omega,y\right), \ v=V\left(\omega,y\right), \
w=W\left(\omega,y\right), \ \omega=t-t_0x.\ee
 So, the reduced system is
\begin{equation}\label{3-46}\begin{array}{l}
d_1\left(t_0^2\,U_{\omega\omega}+U_{yy}\right)-U_{\omega}-UV=0, \medskip\\
d_2\left(t_0^2\,V_{\omega\omega}+V_{yy}\right)-V_{\omega}-UV=0, \medskip\\
d_3\left(t_0^2\,W_{\omega\omega}+W_{yy}\right)-W_{\omega}+UV=0.
\end{array}\end{equation}

The third  algebra from (\ref{3-43}) produces  the ansatz
\be\nonumber u=U\left(\omega_1,\omega_2\right), \
v=V\left(\omega_1,\omega_2\right), \
w=W\left(\omega_1,\omega_2\right), \ \omega_1=x^2+y^2, \
\omega_2=t+t_0\arctan\frac{y}{x},\ee
 and the reduced system
 \begin{equation}\nonumber\begin{array}{l}
4d_1\omega_1U_{\omega_1\omega_1}+\frac{d_1t_0^2}{\omega_1}U_{\omega_2\omega_2}+
4d_1U_{\omega_1}-U_{\omega_2}-UV=0, \medskip\\
4d_2\omega_1V_{\omega_1\omega_1}+\frac{d_2t_0^2}{\omega_1}V_{\omega_2\omega_2}+
4d_2V_{\omega_1}-V_{\omega_2}-UV=0, \medskip\\
4d_3\omega_1UW_{\omega_1\omega_1}+\frac{d_3t_0^2}{\omega_1}W_{\omega_2\omega_2}+
4d_3W_{\omega_1}-W_{\omega_2}+UV=0,
\end{array}\end{equation}
 respectively.

Let us construct exact solutions of system (\ref{3-44}). Using the
Lie symmetry operator $\alpha_2\p_x-\alpha_1\p_y$
($\alpha_1^2+\alpha_2^2\neq0$) of system (\ref{3-44}), one arrives
at the plane wave ansatz \be\label{3-49} u=U\left(\omega\right), \
v=V\left(\omega\right), \ w=W\left(\omega\right), \
\omega=\alpha_1x+\alpha_2y,\ee and the corresponding ODE system
 \begin{equation}\label{3-50}
d_1^*U''=UV, \quad d_2^*V''=UV, \quad d_3^*W''=-UV,
\end{equation} where
$d_i^*=d_i\left(\alpha_1^2+\alpha_2^2\right)$.

Linear combinations of the equations in system (\ref{3-50}) lead to
the equivalent system
 \begin{equation}\label{3-51}\begin{array}{l}
d_1^*U''=UV, \medskip\\
\left(d_1^*U-d_2^*V\right)''=0, \medskip\\
\left(d_1^*U+d_3^*W\right)''=0.
\end{array}\end{equation}

From the last two equations of system (\ref{3-51}), we obtain
 \begin{equation}\label{3-52}\begin{array}{l}
d_2^*V=d_1^*U+b_{21}\omega+b_{20}, \medskip\\
d_3^*W=-d_1^*U+b_{31}\omega+b_{30}, \medskip\\
\end{array}\end{equation} where $b_{ij}$ are arbitrary constants.

 Thus, to solve the ODE system (\ref{3-50}), one
needs to integrate the nonlinear ODE \be\label{3-53}
d_1^*d_2^*U''=U\left(d_1^*U+b_{21}\omega+b_{20}\right).\ee

\emph{For $b_{21}=b_{20}=0$}, the general solution of ODE
(\ref{3-53}) is given by
\be\label{3-54}U=6d_2^*\wp\left(\omega+C_1,0,C_2\right),\ee where
$\wp$ denotes the Weierstrass function.

Using formulae (\ref{2-23*}), (\ref{3-49}), (\ref{3-52}) and
(\ref{3-54}), we obtain the solution of system (\ref{2-23})
\begin{equation}\label{3-55}\begin{array}{l}
u(t,x,y)=6d_2\left(\alpha_1^2+\alpha_2^2\right)\wp\bigl[\left(\alpha_1x+\alpha_2y\right)\sin
2t+
\left(\alpha_1y-\alpha_2x\right)\cos 2t+C_1,0,C_2\bigr], \medskip\\
v(t,x,y)=6d_1\left(\alpha_1^2+\alpha_2^2\right)\wp\bigl[\left(\alpha_1x+\alpha_2y\right)\sin
2t+
\left(\alpha_1y-\alpha_2x\right)\cos 2t+C_1,0,C_2\bigr], \medskip\\
w(t,x,y)=\frac{\alpha_1^2+\alpha_2^2}{d_3}\Big[-6d_1d_2\wp\bigl[\left(\alpha_1x+\alpha_2y\right)\sin
2t+ \left(\alpha_1y-\alpha_2x\right)\cos 2t+C_1,0,C_2\bigr]+\\
\hskip2cm C_3\left(\alpha_1x+\alpha_2y\right)\sin 2t+
C_3\left(\alpha_1y-\alpha_2x\right)\cos 2t+C_4\Big],
\end{array}\end{equation} where the
parameters $\alpha$ and $C$ with subscripts are arbitrary constants.
In the case $C_2=0$, the Weierstrass function degenerates into an
elementary function, therefore  the exact  solution (\ref{3-55})
takes the form
\begin{equation}\label{3-56}\begin{array}{l}
u(t,x,y)=6d_2\left(\alpha_1^2+\alpha_2^2\right)\bigl[\left(\alpha_1x+\alpha_2y\right)\sin
2t+
\left(\alpha_1y-\alpha_2x\right)\cos 2t+C_1\bigr]^{-2}, \medskip\\
v(t,x,y)=6d_1\left(\alpha_1^2+\alpha_2^2\right)\bigl[\left(\alpha_1x+\alpha_2y\right)\sin
2t+
\left(\alpha_1y-\alpha_2x\right)\cos 2t+C_1\bigr]^{-2}, \medskip\\
w(t,x,y)=\frac{\alpha_1^2+\alpha_2^2}{d_3}\Big[-6d_1d_2\bigl[\left(\alpha_1x+\alpha_2y\right)\sin
2t+
\left(\alpha_1y-\alpha_2x\right)\cos 2t+C_1\bigr]^{-2}+\\
\hskip2cm C_3\left(\alpha_1x+\alpha_2y\right)\sin 2t+
C_3\left(\alpha_1y-\alpha_2x\right)\cos 2t+C_4\Big].
\end{array}\end{equation}

\begin{figure}[h!]
\begin{center}
\includegraphics[width=7cm]{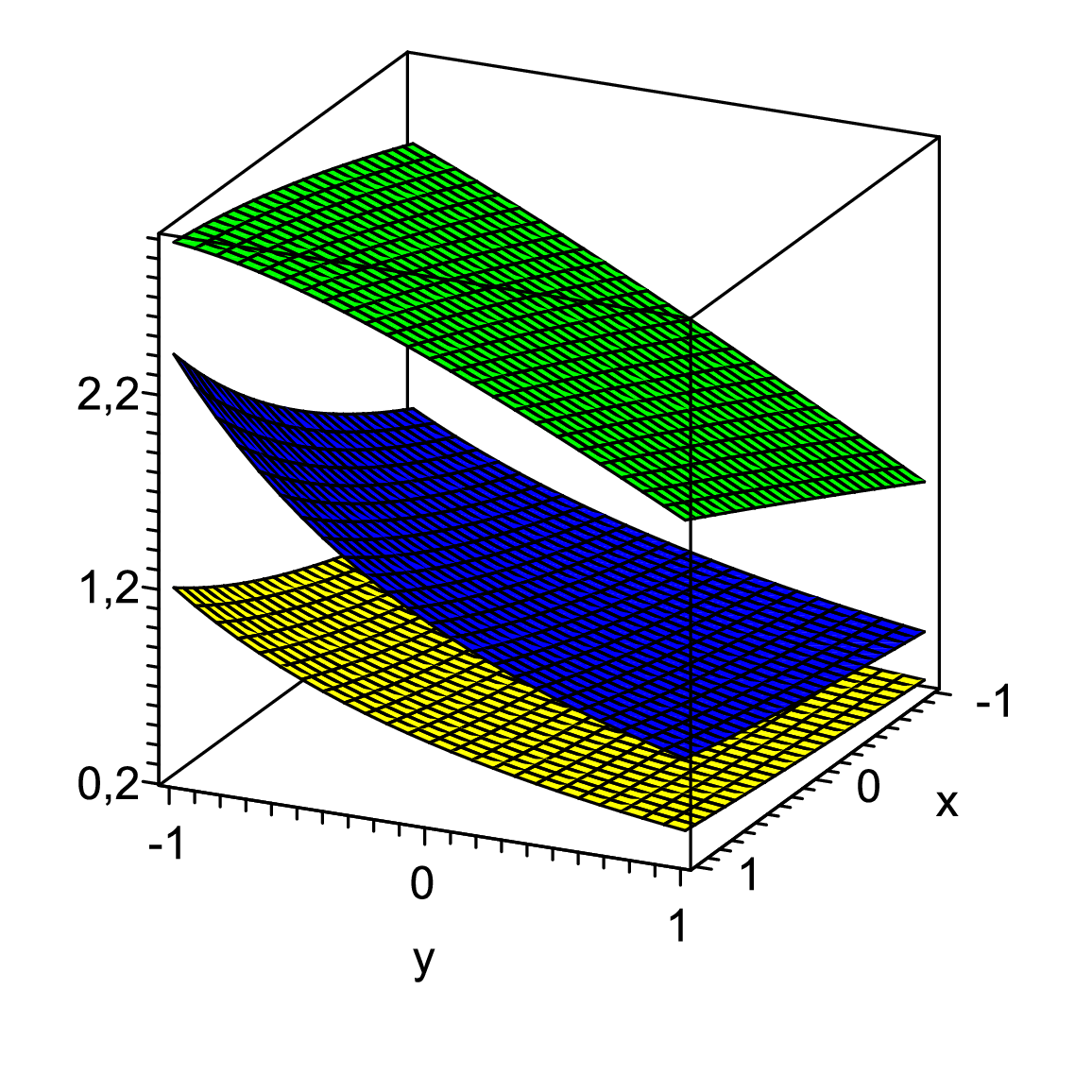}
\includegraphics[width=7cm]{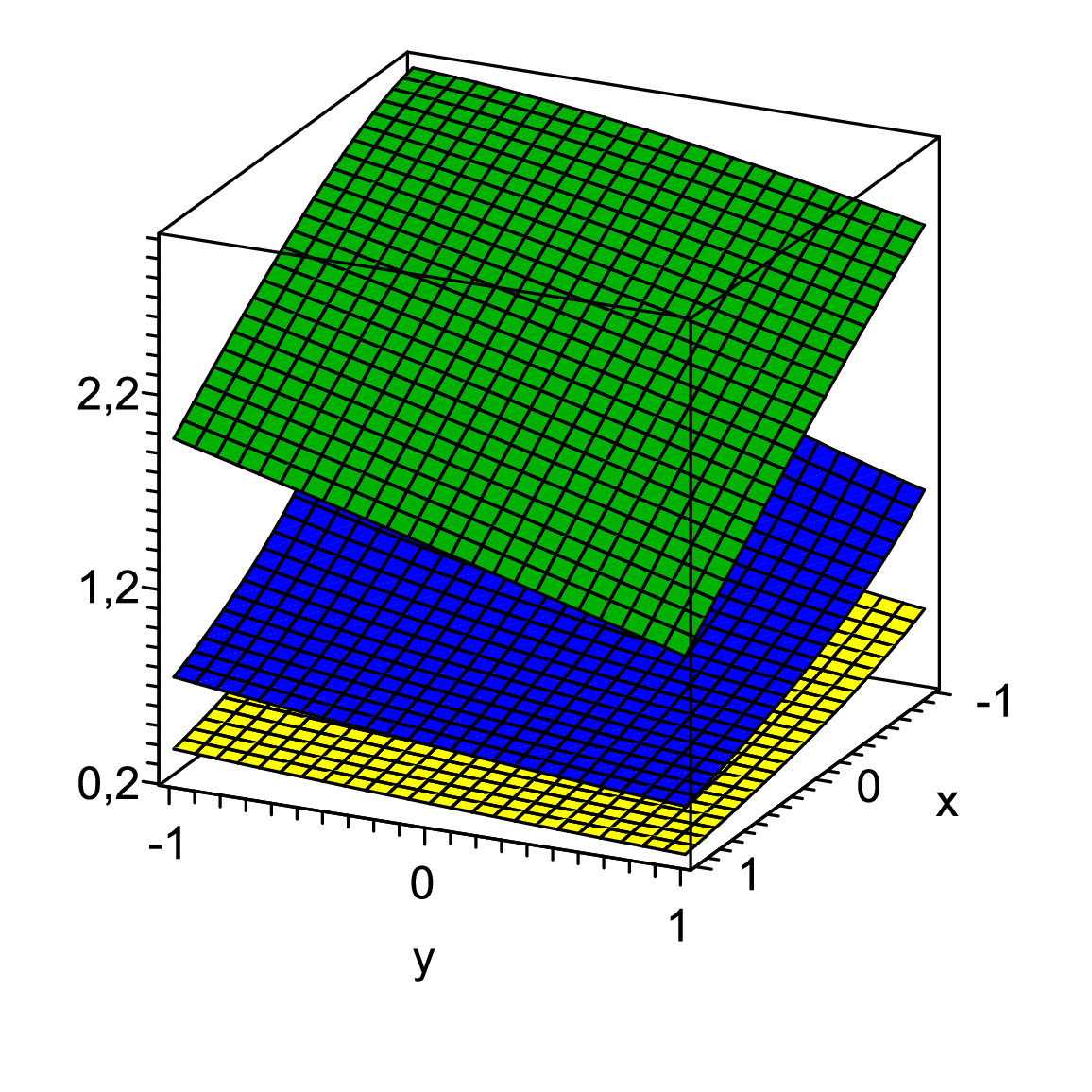}
\includegraphics[width=7cm]{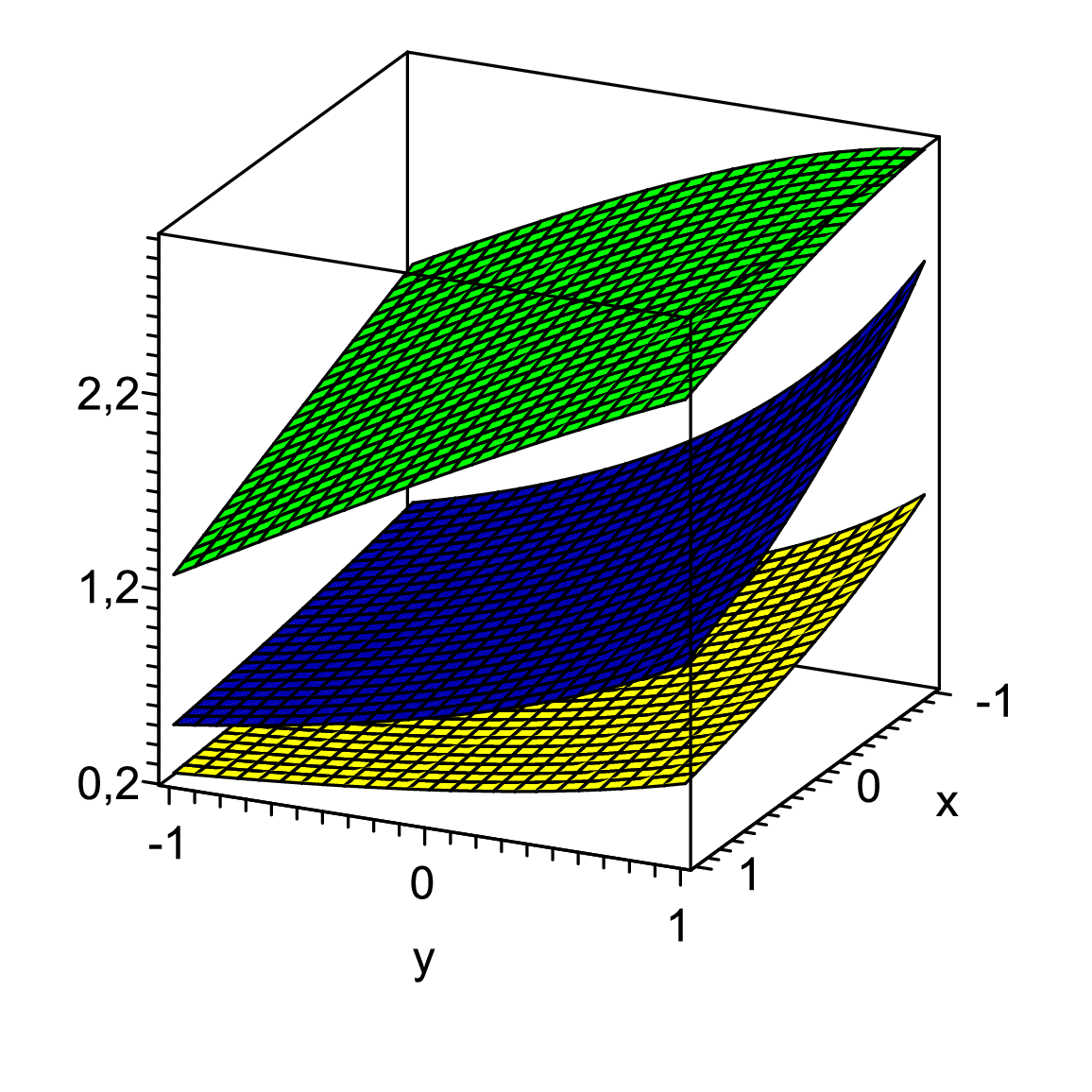}
\includegraphics[width=7cm]{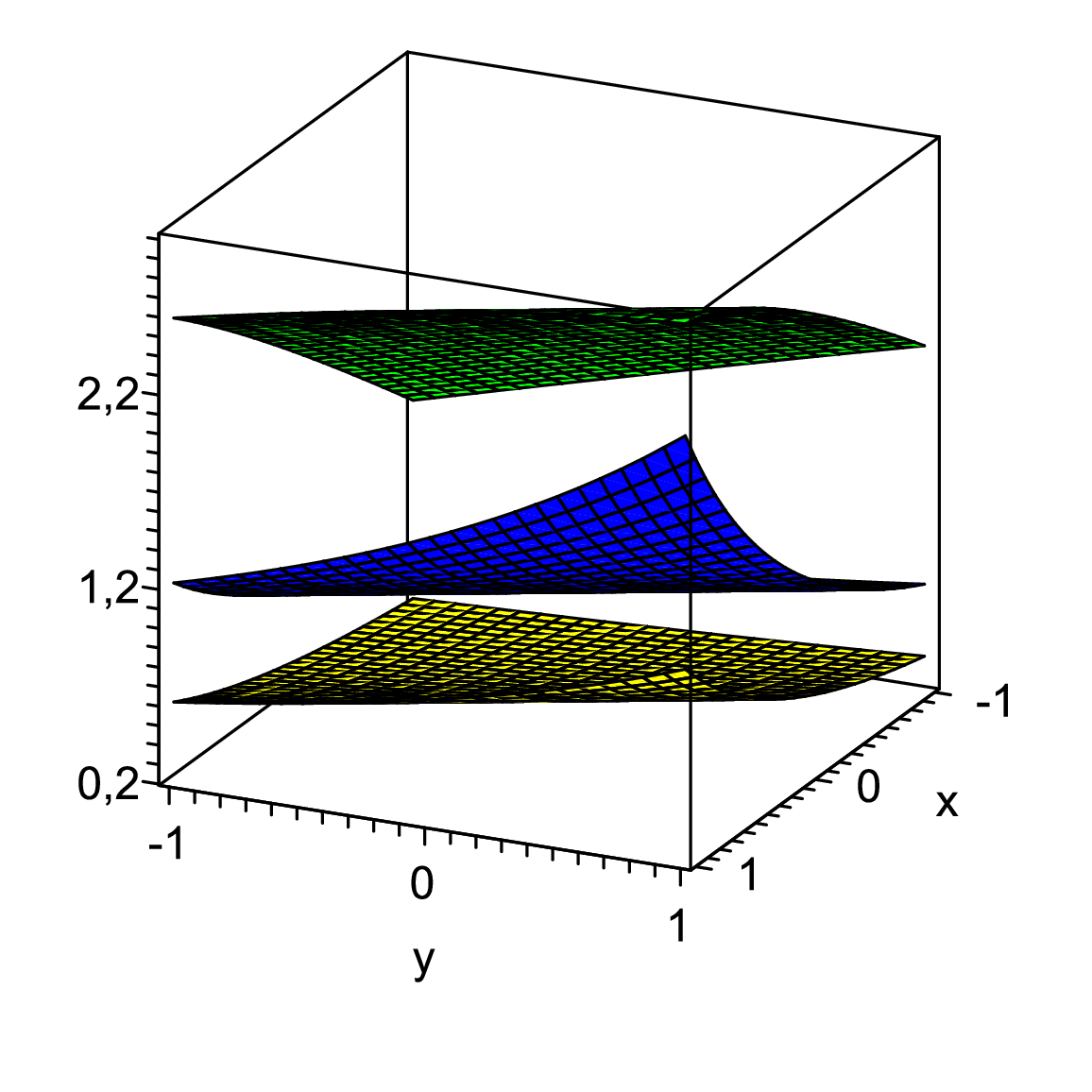}
\end{center}
\caption{Surfaces representing the functions $u(t_0,x,y)$ (blue),
$v(t_0,x,y)$ (yellow) and $w(t_0,x,y)$ (green) from the solution
(\ref{3-56}) of system (\ref{2-23}) with the parameters $d_1=1, \
d_2=2, \ d_3=3, \ \alpha_1=\frac{1}{2}, \ \alpha_2=\frac{1}{4}, \
 C_1=2, \ C_3=-15, \ C_4=55$ and $t_0=0,\frac{\pi}{4},\frac{\pi}{2},\frac{3\pi}{2}$.} \label{f1}
\end{figure}

\begin{figure}[h!]
\begin{center}
\includegraphics[width=7.5cm]{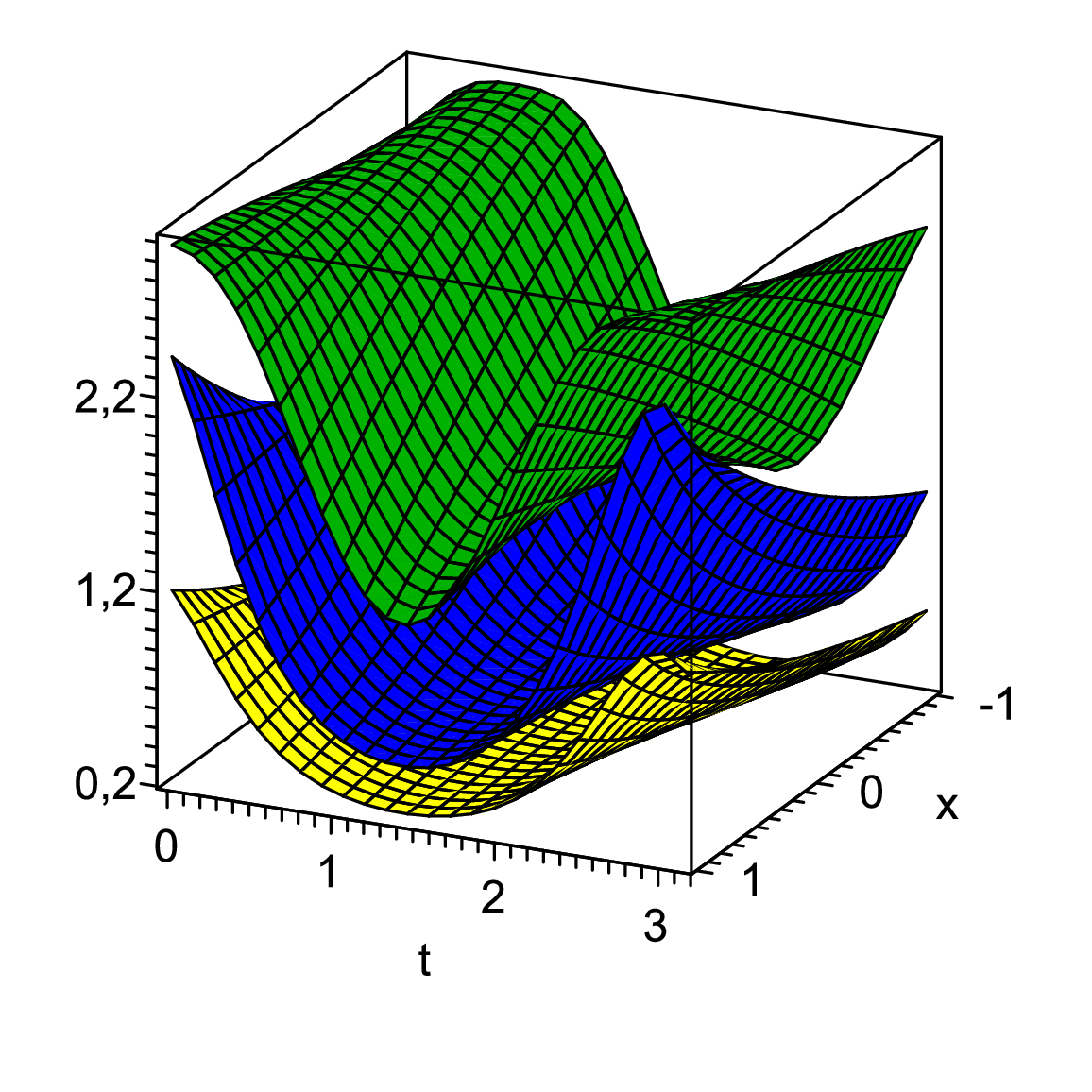}
\includegraphics[width=7.5cm]{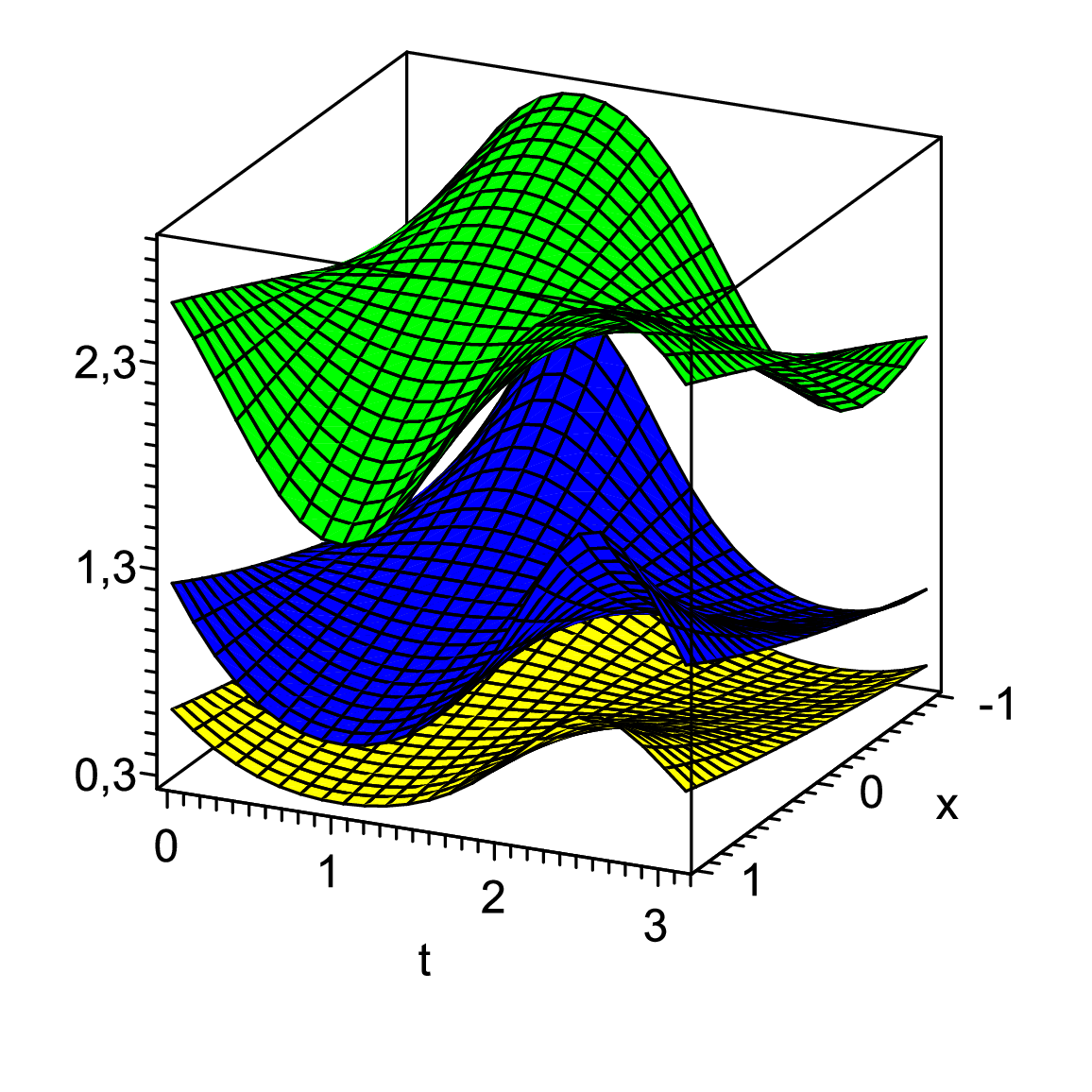}
\includegraphics[width=7.5cm]{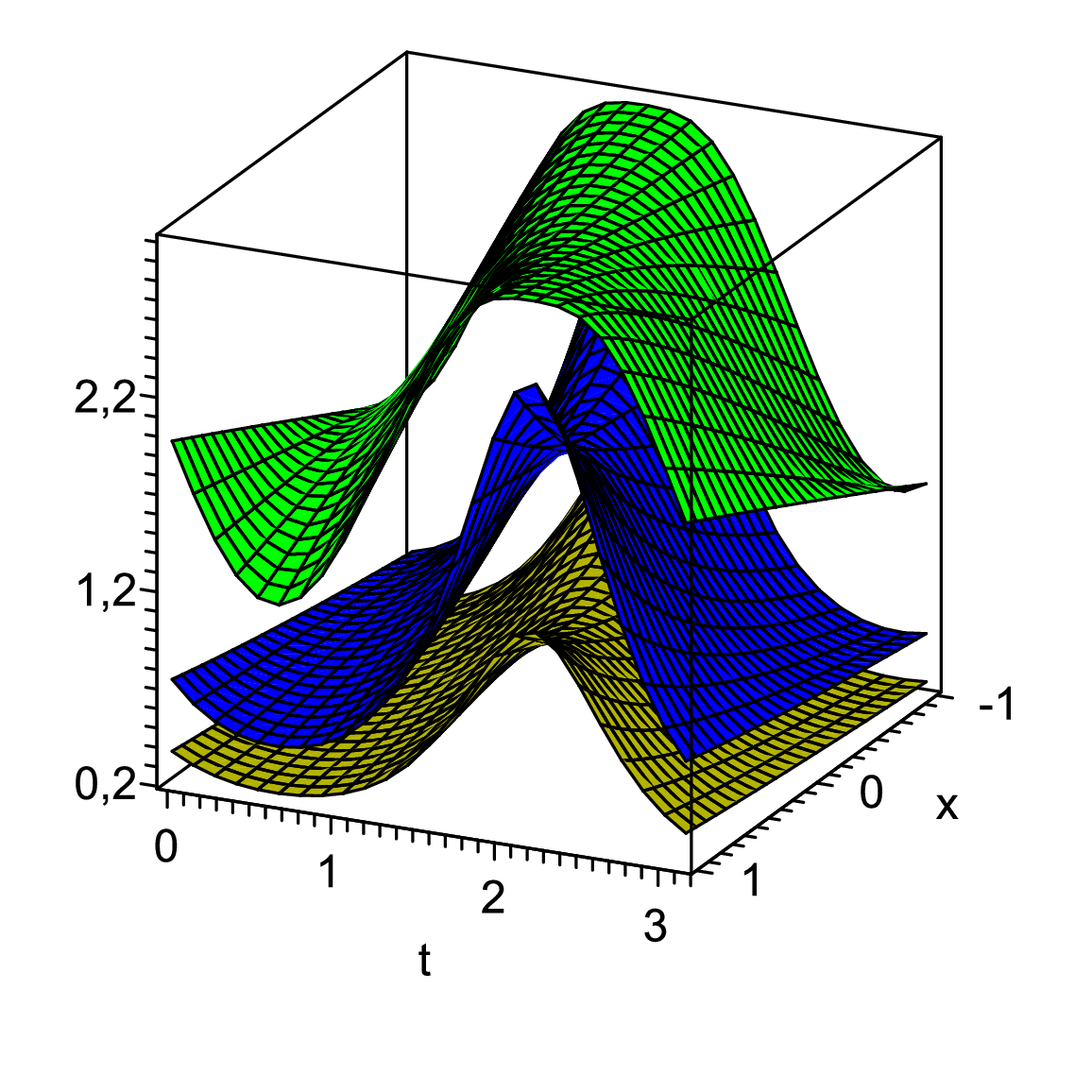}
\end{center}
\caption{Surfaces representing the functions $u(t,x,y_0)$ (blue),
$v(t,x,y_0)$ (yellow) and $w(t,x,y_0)$ (green) from the solution
(\ref{3-56}) of system (\ref{2-23}) with the parameters $d_1=1, \
d_2=2, \ d_3=3, \ \alpha_1=\frac{1}{2}, \ \alpha_2=\frac{1}{4}, \
 C_1=2, \ C_3=-15, \ C_4=55$ and $y_0=-1,0,1$.} \label{f2}
\end{figure}

\begin{figure}[h!]
\begin{center}
\includegraphics[width=7cm]{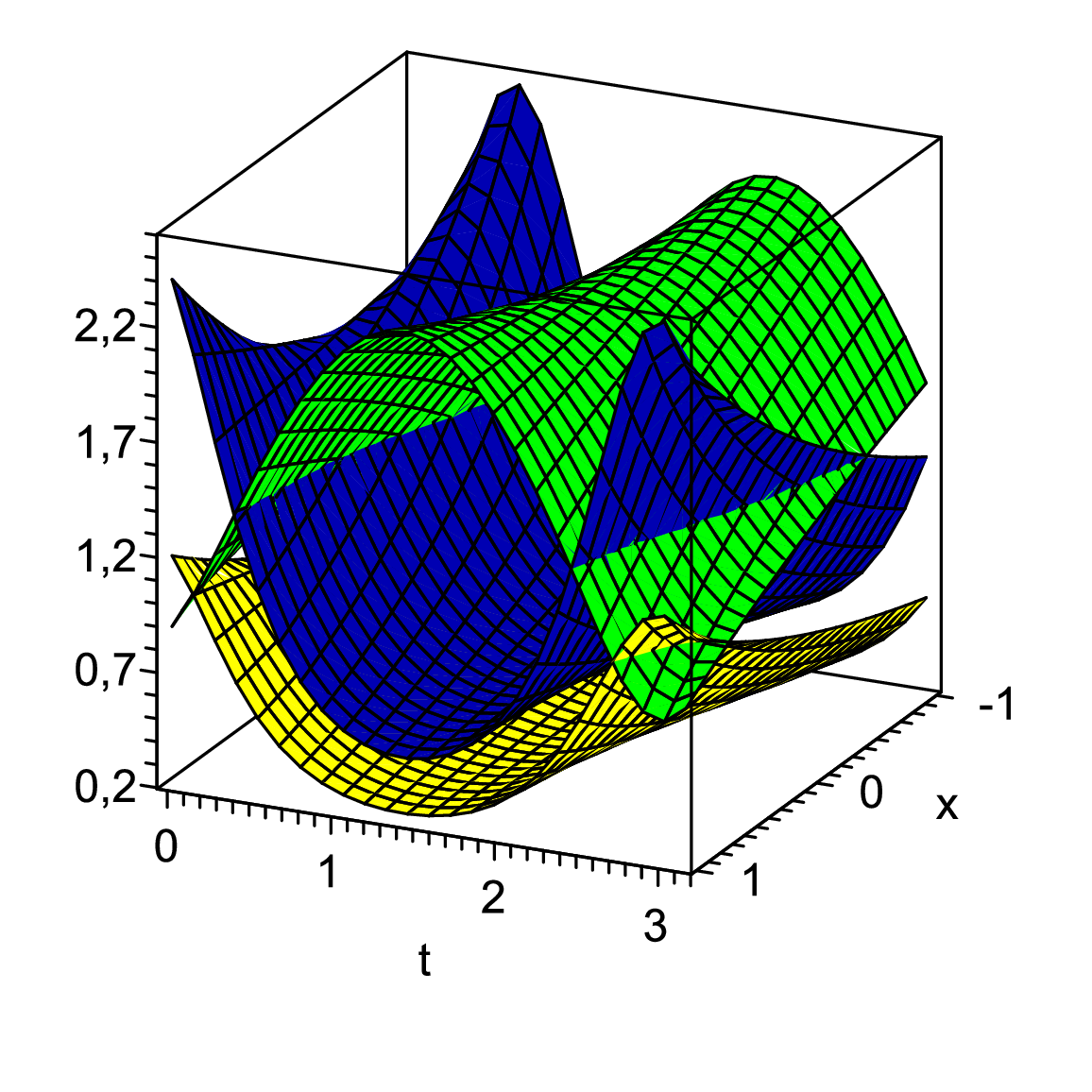}
\includegraphics[width=7cm]{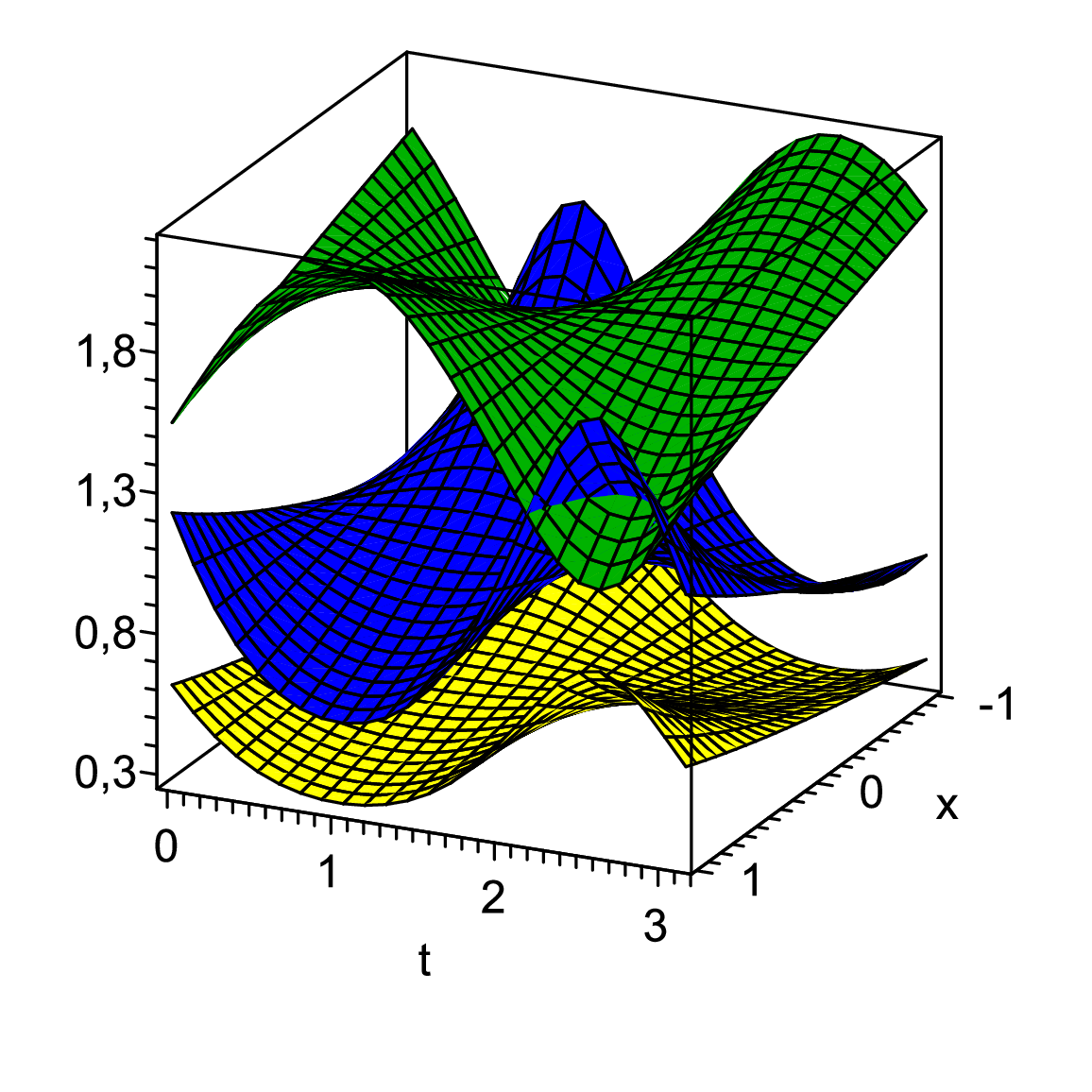}
\includegraphics[width=7cm]{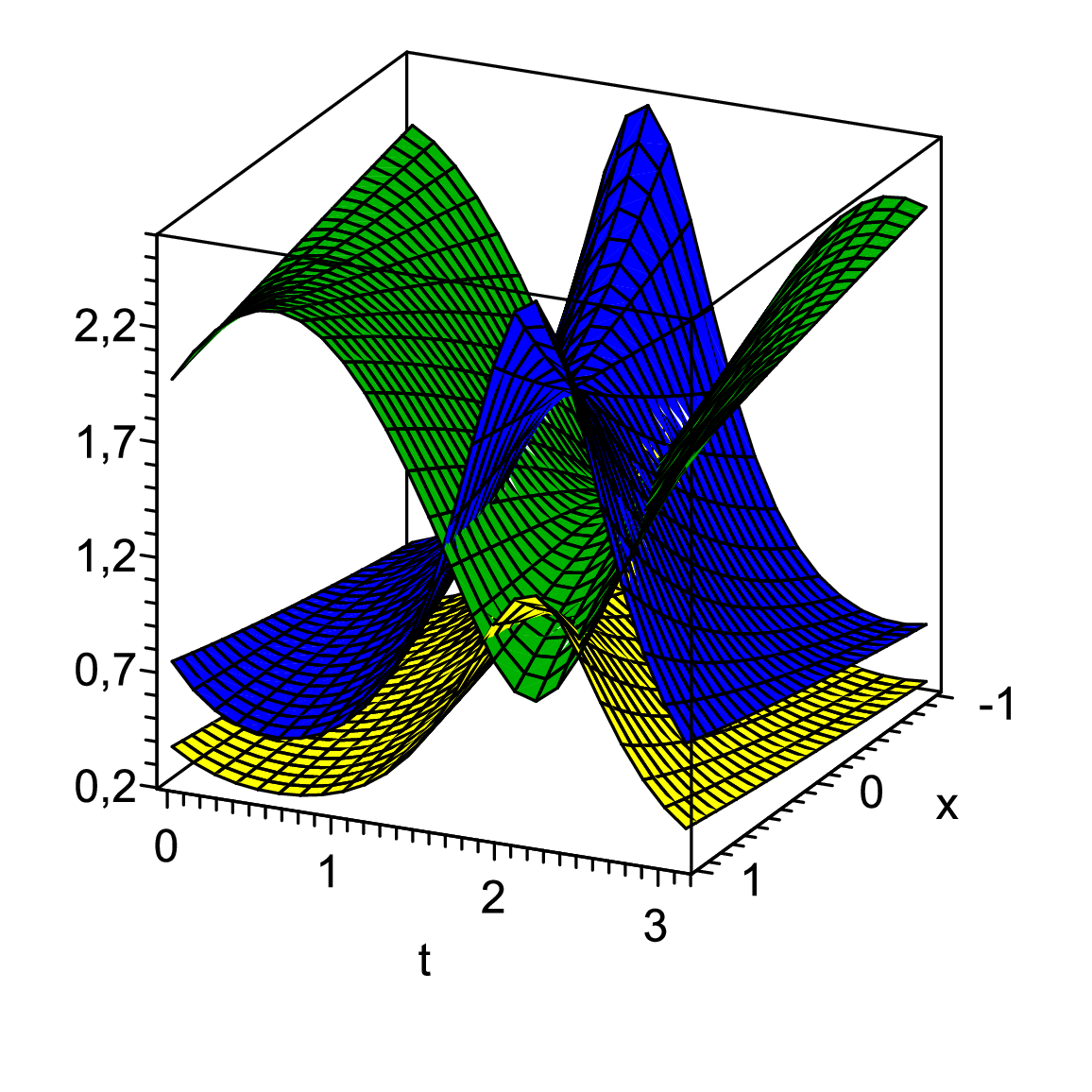}
\end{center}
\caption{Surfaces representing the functions $u(t,x,y_0)$ (blue),
$v(t,x,y_0)$ (yellow) and $w(t,x,y_0)$ (green) from the solution
(\ref{3-56}) of system (\ref{2-23}) with the parameters $d_1=1, \
d_2=2, \ d_3=3, \ \alpha_1=\frac{1}{2}, \ \alpha_2=\frac{1}{4}, \
 C_1=2, \ C_3=5, \ C_4=10$ and $y_0=-1,0,1$.} \label{f3}
\end{figure}

 Formulae (\ref{3-56}) with correctly specified
parameters produce exact solutions of the DLVS with convection
(\ref{2-23}) with positive components. In Fig.~\ref{f1}, \ref{f2}
and~\ref{f3}, two examples are presented. We consider the space
domain $\Omega=(-1,1)\times(-1,1)$ and the time interval $[0,\pi]$
because formulae  (\ref{3-56}) produce periodic solutions. As it
follows from Fig.~\ref{f1} and  \ref{f2}, the concentration $w$
corresponding to the product $C$ of the  chemical   reaction $A+B
\to C$ is higher than that of each reactant in $\Omega$
independently of time. However, such a behaviour of the
concentrations $u, \ v$ and $w$  depends essentially on the
parameters $C_1, \ C_3$ and $C_4$. Fig.~\ref{f3} presents the
concentrations of chemicals for another set of parameters. One
easily notes that the concentration of the product $C$ depending on
time and space can be higher but can be smaller than the
concentrations of the reactants $A$ and $B$. We shall not explore
further the applicability of the exact solution (\ref{3-56}) because
it lies beyond scopes of this study.

\emph{For $b_{21}=0$ and $b_{20}\neq0$}, the general solution of ODE
(\ref{3-53}) can be expressed in terms of elliptic functions. The
relevant formulae are very cumbersome, therefore those are omitted.
The general solution reduces to elementary functions for special
values of the parameter $b_{20}$. For example, setting
$b_{20}=4d_1^*d_2^*C_2^2$, one obtains the function
\[U=-6C_2^2d_2^*\operatorname{sech}^2\left(C_1+C_2\omega\right),\] which is not meaningful
 from the viewpoint of possible interpretation. Another
possible case, $b_{20}=-4d_1^*d_2^*C_2^2$, leads to the solution
\[U=6C_2^2d_2^*\sec^2\left(C_1+C_2\omega\right),\] where $C_1$ and
$C_2$ are arbitrary constants.

Using the above formulae, we obtain the solution of the PDE system
(\ref{2-23})\,:
\begin{equation}\nonumber\begin{array}{l}
u(t,x,y)=6d_2\left(\alpha_1^2+\alpha_2^2\right)C_2^2\sec^{2}\bigl[C_1+C_2\left(\alpha_1x+\alpha_2y\right)\sin
2t+
C_2\left(\alpha_1y-\alpha_2x\right)\cos 2t\bigr], \medskip\\
v(t,x,y)=2d_1\left(\alpha_1^2+\alpha_2^2\right)C_2^2\left(-2+3\sec^{2}\bigl[C_1+C_2\left(\alpha_1x+\alpha_2y\right)\sin
2t+
C_2\left(\alpha_1y-\alpha_2x\right)\cos 2t\bigr]\right), \medskip\\
w(t,x,y)=\frac{\alpha_1^2+\alpha_2^2}{d_3}\Big(-6d_1d_2C_2^2\sec^{2}\bigl[C_1+C_2\left(\alpha_1x+\alpha_2y\right)\sin
2t+
C_2\left(\alpha_1y-\alpha_2x\right)\cos 2t\bigr]+\\
\hskip2cm C_3\left(\alpha_1x+\alpha_2y\right)\sin 2t+
C_3\left(\alpha_1y-\alpha_2x\right)\cos 2t+C_4\Big).
\end{array}\end{equation}

 \emph{For $b_{21}\neq0$}, the general solution of
the nonlinear ODE (\ref{3-53}) is unknown. So, only the trivial
solution $U=-\frac{1}{d_1^*}(b_{21}\omega+b_{20})$ was identified.

 The reduced system  (\ref{3-46}) has a similar
structure to (\ref{3-44}).  So, using the plane wave ansatz
(\ref{3-49}) with $\omega \to z=\alpha_1\omega+\alpha_2y$, i.e.
\[U(\omega,y)=U(z), \quad V(\omega,y)=V(z),
\quad W(\omega,y)=W(z),\] system (\ref{3-46}) reduces to the ODE
system
\begin{equation}\nonumber
d_1^*U''-\alpha_1U'=UV, \quad d_2^*V''-\alpha_1V'=UV, \quad
d_3^*W''-\alpha_1W'=-UV,
\end{equation} where
$d_i^*=d_i\left(t_0^2\alpha_1^2+\alpha_2^2\right)$. In contrast to
the ODE system (\ref{3-50}), application of the same algorithm to
the above system results in the restriction $d_1 = d_2 = d_3 = d$.
Having this restriction in place, the system  can be reduced to
solving the  single ODE
\be\label{3-53*}
d^*U''-\alpha_1U'=U\left(U-b_{1}\exp\left(\frac{\alpha_1}{d^*}z\right)-b_{0}\right),\ee
where $b_{1}$ and $b_{0}$ are arbitrary constants and
$d^*=d\left(t_0^2\alpha_1^2+\alpha_2^2\right)$.

In the case $b_{1}\neq0$, the general solution of the nonlinear ODE
(\ref{3-53*}) is unknown. Only a particular solution of this
equation, $U=b_{1}\exp\left(\frac{\alpha_1}{d^*}z\right)+b_{0}$, can
be found, which leads to the trivial result $V=0.$

Equation (\ref{3-53*}) with $b_{1}=0$ corresponds to the known ODE
that arises when one seeks for travelling waves  of the famous
Fisher equation \cite{fi-37}
\be\label{3-53**}u_{t}=d^*u_{xx}+u(b_{0}-u),\ee using the ansatz
$u=U(z), \ z=x+\alpha_1t.$ The well-known solution of the Fisher
equation (\ref{3-53**})  was identified for the first time in
\cite{abl-zep} and has the form \be\label{3-53a}
u(t,x)=b_0\left(1+C\exp\left[\sqrt{\frac{b_0}{6d^*}}\left(x-5\sqrt{\frac{b_0d^*}{6}}t\right)\right]\right)^{-2},\ee
where $C$ is an arbitrary constants. Solution (\ref{3-53a}) can be
reduced to the form \be\nonumber
u(t,x)=\frac{b_0}{4}\left(1-\tanh\left[\sqrt{\frac{b_0}{24d^*}}\left(x-5\sqrt{\frac{b_0d^*}{6}}t\right)\right]\right)^{2},\ee
in the case $C>0$, and to the form \be\nonumber
u(t,x)=\frac{b_0}{4}\left(1-\coth\left[\sqrt{\frac{b_0}{24d^*}}\left(x-5\sqrt{\frac{b_0d^*}{6}}t\right)\right]\right)^{2},\ee
in the case $C<0$.

Using the above formulae, we obtain the exact solutions of the DLVS
with convection   (\ref{2-23}) with $d_1 = d_2 = d_3 = d$ in the
following forms\,:
\begin{equation}\nonumber\begin{array}{l}
u(t,x,y)=\frac{3C_1}{5}\Bigl[1+\tanh\bigl[C_1t+\left(\alpha_1x+\alpha_2y\right)\sin
2t+\left(\alpha_1y-\alpha_2x\right)\cos 2t\bigr]\Bigr]^{2}, \medskip\\
v(t,x,y)=-\frac{12C_1}{5}+u, \medskip\\
w(t,x,y)=C_2+C_3\exp\Big[10C_1t+10\left(\alpha_1x+\alpha_2y\right)\sin
2t+10\left(\alpha_1y-\alpha_2x\right)\cos 2t\Big]-u,
\end{array}\end{equation} and
\begin{equation}\nonumber\begin{array}{l}
u(t,x,y)=\frac{3C_1}{5}\Bigl[1+\coth\bigl[C_1t+\left(\alpha_1x+\alpha_2y\right)\sin
2t+\left(\alpha_1y-\alpha_2x\right)\cos 2t\bigr]\Bigr]^{2}, \medskip\\
v(t,x,y)=-\frac{12C_1}{5}+u, \medskip\\
w(t,x,y)=C_2+C_3\exp\Big[10C_1t+10\left(\alpha_1x+\alpha_2y\right)\sin
2t+10\left(\alpha_1y-\alpha_2x\right)\cos 2t\Big]-u,
\end{array}\end{equation}
 where $\alpha_1, \ \alpha_2, \ C_2$ and $C_3$
are arbitrary constants, while
$C_1=10d\left(\alpha_1^2+\alpha_2^2\right).$

\begin{remark}
Equation (\ref{3-53*}) with $b_{1}=0$ possesses also exact solutions
involving the Weierstrass function provided its parameters satisfy
some algebraic restrictions reducing this ODE to that listed in 6.23
\cite{kamke}. So, exact solutions of system (\ref{2-23}) with $d_1 =
d_2 = d_3 = d$ involving the Weierstrass function can be
constructed.
\end{remark}

\subsection{Exact solutions of the  diffusive Lotka--Volterra type system   with an arbitrary  radially-symmetric stream function }
\label{sec-3.1} Let us construct exact solutions of system
(\ref{2-1}) when the latter  involves an arbitrary  function $\Psi$
as presented in Case 1 of Table~\ref{tab1}. Setting $\beta=0$ for
simplicity and $k=1$ (without loss of generality), system
(\ref{2-1}) takes the form
\begin{equation}\label{3-1}\begin{array}{l}
u_t-\frac{\alpha\left(xu_x+yu_y\right)}{x^2+y^2}+2F'\left(xu_y-yu_x\right) = d_1\left(u_{xx}+u_{yy}\right)-uv, \medskip\\
v_t-\frac{\alpha\left(xu_x+yu_y\right)}{x^2+y^2}+2F'\left(xu_y-yu_x\right) = d_2\left(v_{xx}+v_{yy}\right)-uv, \medskip\\
w_t-\frac{\alpha\left(xu_x+yu_y\right)}{x^2+y^2}+2F'\left(xu_y-yu_x\right) = d_3\left(w_{xx}+w_{yy}\right)+uv, \medskip\\
\end{array}\end{equation}
where $F'=\frac{dF}{d(r^2)}, \ r^2=x^2+y^2$.

Since system (\ref{3-1}) admits the rotation operator
\[ x \p_y -y\p_x,\]
it is reducible to the $(1+1)$-dimensional system
\begin{equation}\label{3-5}\begin{array}{l}
u_t = d_1u_{rr}+\frac{d_1+\alpha}{r}\,u_r-uv, \medskip\\
v_t=d_2v_{rr}+\frac{d_2+\alpha}{r}\,v_r-uv, \medskip\\
w_t=d_3w_{rr}+\frac{d_3+\alpha}{r}\,w_r+uv, \medskip\\
\end{array}\end{equation} by the ansatz
\be\label{3-4}u=u(t,r), \ v=v(t,r), \ w=w(t,r), \
r=\sqrt{x^2+y^2}.\ee

Let us consider the stationary case, i.e. all unknown
  functions are independent of the time variable.
  In this case, the nonlinear PDE system (\ref{3-5}) reduces to a system of nonlinear second-order ODEs\,:
\begin{equation}\label{3-17}\begin{array}{l}
d_1u''+\frac{d_1+\alpha}{r}\,u'-uv=0, \medskip\\
d_2v''+\frac{d_2+\alpha}{r}\,v'-uv=0, \medskip\\
d_3w''+\frac{d_3+\alpha}{r}\,w'+uv=0, \medskip\\
\end{array}\end{equation}

As a first step, we determine the functions $u$ and $v$ from the
first two equations of system (\ref{3-17}). Depending on the values
of the parameters $d_1, \ d_2$ and $\alpha$, we obtain the following
three cases.

1. When the parameters $d_1, \ d_2$ and $\alpha$ are arbitrary
constants, the functions $u(r)$ and $v(r)$ are given by
\be\label{3-28}u(r)=d_2C+\alpha f(r)+d_2rf'(r), \quad
v(r)=d_1C+\alpha f(r)+d_1rf'(r),\ee where $f(r)$ is a solution of
the nonlinear equation
\be\label{3-18}\ba d_1d_2r^2f'''+(3d_1d_2+d_1\alpha+d_2\alpha)rf''-d_1d_2r^3{f'}^2+\Big(-\alpha(d_1+d_2)r^2f-2Cd_1d_2r^2+\medskip\\
\hskip2cm (\alpha+d_1)(\alpha+d_2)\Big)f'-\alpha^2rf^2-\alpha
C(d_1+d_2)rf-C^2d_1d_2r=0.\ea\ee Hereafter $C$ (with or without
subscripts) denotes an arbitrary constant.

It is extremely difficult to find any solutions of the nonlinear
third-order ODE (\ref{3-18}) in the general case. Additional
assumptions must be applied in order to determine the function~$f$.
Assuming that the function $f$ is of the following form
$f=f_0r^{\beta}$ (here $f_0\neq0$ and $\beta\neq0$ are arbitrary
constants), one obtains a solution of equation (\ref{3-18}), namely
$f=-2r^{-2}$, under the restriction $C=0$. Taking into account
(\ref{3-28}) and integrating the third equation of system
(\ref{3-17}), we arrive at the solution \be\label{3-21}\ba
u(r)=2(2d_2-\alpha)\,r^{-2}, \medskip\\
v(r)=2(2d_1-\alpha)\,r^{-2},\medskip\\
w(r)=\left\{
\begin{array}{l}C_0+C_1r^{-\frac{\alpha}{d_3}}+\frac{2(2d_1-\alpha)(2d_2-\alpha)}{\alpha-2d_3}\,r^{-2}, \ \alpha(\alpha-2d_3)\neq0,\medskip\\
C_0+C_1r^{-2}+\frac{4(d_1-d_3)(d_2-d_3)}{d_3r^2}\,(1+2\ln r), \ \alpha=2d_3,\medskip\\
C_0+C_1\ln r-\frac{4d_1d_2}{d_3r^2}, \ \alpha=0,
  \end{array} \right.\ea\ee of the ODE system (\ref{3-17}).

2. In the case $\alpha=0$, the following expressions can  easily be
derived from (\ref{3-17}) \[u(r)=d_2f(r)+C_1\ln r-C_0, \quad
v(r)=d_1f(r),\] where $f(r)$ is a solution of the nonlinear equation
\be\label{3-19} rf''+f'-rf^2-\frac{C_1\ln r-C_0}{d_2}\,rf=0.\ee

 To the best of our knowledge,  the nonlinear ODE (\ref{3-19}) is not integrable.
Notably, in the case $C_1=0$, the above ODE takes the form
\be\label{3-19*} rf''+f'-rf^2+\frac{C_0}{d_2}\,rf=0, \ee
which belongs to the class of Emden–Fowler type equations.  ODE
(\ref{3-19*}) with  $C_0=0$ is a particular case of the modified
Emden–Fowler equation. Although some particular cases of this
equation are integrable (see, e.g., \cite{pol-za-2018}), the general
solution of ODE (\ref{3-19*}) is unknown. We were able to identify
only a particular solution  in the form $f=4r^{-2}$, which yields
the already obtained solution (\ref{3-21}) with $\alpha=0$.

3. In the case $\alpha\neq0$ and $d_1=d_2=d$ (one can set $d=1$
without loss of generality)
\be\label{3-25}u(r)=f(r)+C_1r^{-\alpha}+C_0, \quad v(r)=f(r),\ee
where  $f(r)$ is a solution of the nonlinear equation
\be\label{3-20}rf''+(1+\alpha)f'-rf^2-\left(C_1r^{-\alpha}+C_0\right)rf=0.\ee
 Setting $\alpha=-1$ and $C_1=0$ for simplicity, the
general solution of equation (\ref{3-20}) can be constructed in the
form \be\nonumber
\pm\int\frac{df}{\sqrt{\frac{2}{3}f^3+C_0f^2+C_2}}=r+C_3.\ee The
above integral leads to elliptic functions provided $C_0$ and $ C_2$
are arbitrary. However, there are several cases when
 exact solutions of the ODE (\ref{3-20}) are obtainable in terms of
elementary functions. Setting, for example, $C_2=-\frac{C_0^3}{3}$
and $C_3=0$, one arrives at the solution of the ODE~(\ref{3-20})
\be\label{3-22**} f(r)=\left\{
\begin{array}{l} 2\beta^2\left(3\sec^2(\beta r)-2\right), \
C_0=4\beta^2,
 \medskip \\ 2\beta^2\left(2-3\operatorname{sech}^2(\beta r)\right), \
C_0=-4\beta^2,
\end{array} \right.    \ee
where $\beta\neq0$ is an arbitrary constant.

Taking into account (\ref{3-25}) and the first expression for the
function $f(r)$ from  (\ref{3-22**}), and integrating the third
equation of system (\ref{3-17}) with $d_3=1$ (for simplicity), we
obtain the following solution\,: \be\label{3-23}\ba
u(r)=6\beta^2\sec^2(\beta r), \medskip\\
v(r)=2\beta^2\left(3\sec^2(\beta r)-2\right), \medskip\\
w(r)=C_4+C_5r-6\beta^2\sec^2(\beta r).\ea\ee


Thus, using formulae (\ref{3-4}), (\ref{3-21}) and (\ref{3-23}), we
obtain the following  steady-state solutions: \be\nonumber \ba
u=\frac{2(2d_2-\alpha)}{x^2+y^2}, \medskip\\
v=\frac{2(2d_1-\alpha)}{x^2+y^2},\medskip\\
w=\left\{
\begin{array}{l}C_0+C_1\left(x^2+y^2\right)^{-\frac{\alpha}{2d_3}}+\frac{2(2d_1-\alpha)(2d_2-\alpha)}{(\alpha-2d_3)\left(x^2+y^2\right)}, \ \alpha(\alpha-2d_3)\neq0,\medskip\\
C_0+\frac{C_1}{x^2+y^2}+\frac{4(d_1-d_3)(d_2-d_3)}{d_3\left(x^2+y^2\right)}\,\left(1+\ln \left(x^2+y^2\right)\right), \ \alpha=2d_3,\medskip\\
C_0+C_1\ln
\left(x^2+y^2\right)-\frac{4d_1d_2}{d_3\left(x^2+y^2\right)}, \
\alpha=0,
  \end{array} \right.\ea\ee
of system (\ref{3-1}) with arbitrary diffusivities, and
\be\nonumber\ba
u=6\beta^2\sec^2\left(\beta \sqrt{x^2+y^2}\right), \medskip\\
v=2\beta^2\left(3\sec^2\left(\beta \sqrt{x^2+y^2}\right)-2\right), \medskip\\
w=C_4+C_5\sqrt{x^2+y^2}-6\beta^2\sec^2\left(\beta
\sqrt{x^2+y^2}\right),\ea\ee of system (\ref{3-1}) with
$d_1=d_2=d_3=1$ and $\alpha=-1$.

 Now we return  to the nonlinear system (\ref{3-5})
and our aim is to construct nonstationary solutions. It can be
verified that system (\ref{3-5}) admits the Lie symmetry
\be\nonumber 2t\p_t+r\p_r-2u\p_u-2v\p_v-2w\p_w,\ee which leads to
the ansatz \be\label{3-7} u=\frac{U\left(\omega\right)}{t}, \
v=\frac{V\left(\omega\right)}{t}, \
w=\frac{W\left(\omega\right)}{t}, \ \omega=\frac{r^2}{t},\ee where
$U, \ V$ and $W$ are new unknown functions. Substituting ansatz
(\ref{3-7}) into  (\ref{3-5}), we obtain the ODE system\,:
\begin{equation}\label{3-7*}\begin{array}{l}
4d_1\,\omega\, U''+\left(4d_1+2\alpha+\omega\right)U'+U\left(1-V\right)=0, \medskip\\
4d_2\,\omega\, V''+\left(4d_2+2\alpha+\omega\right)V'+V\left(1-U\right)=0, \medskip\\
4d_3\,\omega\, W''+\left(4d_3+2\alpha+\omega\right)W'+W+UV=0,
\end{array}\end{equation}
where $U''=\frac{d^2U}{d\omega^2}, \ U'=\frac{dU}{d\omega}, \dots$ .

 The first two equations of system (\ref{3-7*}) form the stationary Lotka--Volterra system in the radially-symmetric case.
To the best of our knowledge, its nontrivial solutions are unknown.
In order to find examples of
 exact solutions, we use ad hoc ansatz
\be\nonumber U=c_{11}\omega^\nu+c_{10}, \quad
V=c_{21}\omega^\nu+c_{20} \ee where $c_{ij}$ and $\nu\neq0$ are
to-be-determined  constants. Substituting the above ansatz into the
first two equations of  (\ref{3-7*})  and  making standard routine,
one obtains\,: \be\label{3-10}U=2(2d_2-\alpha)\,\omega^{-1}, \
V=2(2d_1-\alpha)\,\omega^{-1},\ee if  $c_{10}=c_{20}=0$, and
\be\label{3-13}U=2(d_2-d_1)\,\omega^{-1}+1, \
V=2(d_1-d_2)\,\omega^{-1}+1, \ \alpha=d_1+d_2,\ee if
$c_{10}=c_{20}=1$. Substituting the functions $U$ and $V$ from
(\ref{3-10}) into the third equation of the ODE system (\ref{3-7*}),
one easily derives  the function $W$ in the form
\be\label{3-11}W=\omega^{-\frac{\alpha}{2d_3}}e^{-\frac{\omega}{4d_3}}\left(C_2+
\frac{1}{d_3}\int\omega^{\frac{\alpha}{2d_3}-2}e^{\frac{\omega}{4d_3}}\left((2d_1-\alpha)(2d_2-\alpha)+C_1\omega\right)d\omega\right).\ee
The integral on the right-hand side of (\ref{3-11}) can be evaluated
explicitly in terms of elementary functions only for certain values
of
 $\alpha$ and $C_1$.
Probably the most general case occurs when $\alpha=2nd_3 \
(n=2,3,4,\dots)$ and  $C_1$ is an arbitrary constant. As a result,
one obtains $W$ in the form \be\label{3-11*}\ba
W=\frac{4C_1}{\omega}+C_2\omega^{-n}e^{-\frac{\omega}{4d_3}}+\medskip \\
16\Big(\left(d_1-nd_3\right)\left(d_2-nd_3\right)-(n-1)d_3C_1\Big)\omega^{-n}\left(\sum\limits_{k=0}^{n-2}
(-1)^k \frac{(n-2)!}{(n-2-k)!} (4 d_3)^{n-2-k} \omega^k\right).
\ea\ee The simplest solution arises when $\alpha=0$ and
$C_1=\frac{d_1d_2}{d_3}$: \be \label{3-29} W=
C_2e^{-\frac{\omega}{4d_3}}-\frac{4d_1d_2}{d_3\omega}. \ee

Similarly, substituting the functions $U$ and $V$ from (\ref{3-13})
into the third equation of the ODE system (\ref{3-7*}), we obtain
\be\nonumber
W=\omega^{-\frac{d_1+d_2}{2d_3}}e^{-\frac{\omega}{4d_3}}\left(C_2+
\frac{1}{4d_3}\int\omega^{\frac{d_1+d_2}{2d_3}-1}e^{\frac{\omega}{4d_3}}\left(C_1-\omega-\frac{4(d_1-d_2)^2}{\omega}\right)d\omega\right).\ee
Setting $\alpha=d_1+d_2=2nd_3, \ n=2,3,4\dots $, we again derive $W$
in the form that is quite similar to  (\ref{3-11*}). There is also
the solution \be\nonumber W=
C_2\omega^{-\frac{d_1+d_2}{2d_3}}e^{-\frac{\omega}{4d_3}}-\frac{2\left(d_1-d_2\right)^2}{\left(d_1+d_2-2d_3\right)\omega}-1,
\ee if
$C_1=\frac{4\left(d_1^2+d_2^2-d_1d_3-d_2d_3\right)}{2d_3-d_1-d_2}$
and $ 2d_3\neq d_1+d_2$.

 Thus, taking into account formulae (\ref{3-4}), (\ref{3-7}), (\ref{3-10}),
(\ref{3-13}) and (\ref{3-29}), we obtain the exact solution
\begin{equation}\nonumber\begin{array}{l}
u=\frac{4d_2}{x^2+y^2}, \medskip\\
v=\frac{4d_1}{x^2+y^2},\medskip\\
w=
\frac{C_2}{t}\exp\left(-\frac{x^2+y^2}{4d_3t}\right)-\frac{4d_1d_2}{d_3(x^2+y^2)}
\end{array}\end{equation}
of the nonlinear system
\begin{equation}\nonumber\begin{array}{l}
u_t+2F'\left(xu_y-yu_x\right) = d_1\left(u_{xx}+u_{yy}\right)-uv, \medskip\\
v_t+2F'\left(xu_y-yu_x\right) = d_2\left(v_{xx}+v_{yy}\right)-uv, \medskip\\
w_t+2F'\left(xu_y-yu_x\right) = d_3\left(w_{xx}+w_{yy}\right)+uv.
\end{array}\end{equation}
Similarly, the exact solution
\begin{equation}\nonumber\begin{array}{l}
u=\frac{1}{t}-\frac{2(d_1-d_2)}{x^2+y^2},  \medskip\\
v=\frac{1}{t}+\frac{2(d_1-d_2)}{x^2+y^2}, \medskip\\
w=
C_2\exp\left(-\frac{x^2+y^2}{4d_3t}\right)\,t^{\frac{d_1+d_2-2d_3}{2d_3}}\left(x^2+y^2\right)^{-\frac{d_1+d_2}{2d_3}}-
\frac{2\left(d_1-d_2\right)^2}{\left(d_1+d_2-2d_3\right)\left(x^2+y^2\right)}-\frac{1}{t}
\end{array}
\end{equation}
 of the nonlinear system (\ref{3-1}) with $\alpha=d_1+d_2$ is obtained.
 Notably, using formula (\ref{3-11*}), exact solutions of (\ref{3-1})  with more complicated forms can be written down.

\section{Conclusions} \label{sec-4}

 This work is devoted to the mathematical model
(\ref{0-1})\cite{ge-de_wit-2009}, which has been  introduced for
description of the viscous fingering induced by the chemical
reaction consisting of three components. This model is formed by a
five-component nonlinear system consisting of first- and
second-order PDEs. In order to simplify the model, the
  stream function  was introduced, therefore  the
 five-component system was reduced to a three-component DLV type
system, involving convective terms. A complete Lie symmetry
classification of the three-component system (\ref{2-1}) was
performed. As a result, it was proved that exactly 11 forms of the
stream function arise leading to nontrivial Lie symmetry of
(\ref{2-1}). Any other system possessing a nontrivial Lie symmetry
is reducible to those from Table~\ref{tab1} by one of equivalence
transformations identified in Theorem~\ref{th1}. It was revealed
that the DLVS with convection (\ref{2-1}) with the stream functions
corresponding to constant and linear velocity fields possess the
widest Lie algebras of invariance (see Cases 10 and 11 in
Table~\ref{tab1}). Moreover, nontrivial transformations were
identified that reduce (\ref{2-1}) to that without convective terms.
In the case of a constant velocity field, this transformation is
nothing else but the Galilei boost, however,  the highly nontrivial
transformation (\ref{2-23*}) was found for the linear velocity
field. Transformation (\ref{2-23*}) was identified via a careful
analysis of the Lie algebra of invariance of the relevant
three-component system.

 The DLVS with
convection  corresponding the linear velocity
 field  was examined in detail.
Using its Lie symmetries, several multiparameter families of exact
solutions were constructed. These solutions include time-dependent
and  radially symmetric solutions as well as more complicated
solutions expressed in terms of the Weierstrass function.
It was shown that some of  exact  solutions can be used for
demonstration of  spatiotemporal evolution of concentrations
corresponding to two reactants  and  their product. In Fig.~\ref{f2}
and~\ref{f3},  the plots of concentrations  are presented for
specified sets of parameters.

Finally, we consider system (\ref{3-1}) with  the most general form
of the stream function arising in Case 1 of Table~\ref{tab1}. It is
shown that the function $F$ disappears if one looks for
radially-symmetric solutions. It means that only the part
corresponding to the   flow field $(U_1,U_2)=\Big(\frac{x}{x^2+y^2},
\frac{y}{x^2+y^2}\Big)$  has impact on  radially-symmetric
solutions. By reducing of system (\ref{3-1}) to the standard form of
the DLVS for search for radially-symmetric solutions, we were able
to construct several examples of exact solutions in explicit forms.
The exact  solutions derived  may provide insights into the
qualitative behavior of chemical reaction–diffusion processes
described by the DLVS  with convection (\ref{2-1}) and  the original
model (\ref{0-1}).  They can be used for checking accuracy of
numerical simulations as well.

\section{Acknowledgments} R.Ch.
 acknowledges that this research was partially  funded
by  the British Academy (Leverhulme Researchers at Risk Research
Support Grant LTRSF-24-100101). This research was partially
supported by a grant from the Simons Foundation
(SFI-PD-Ukraine-00014586, V.D.).

\end{document}